\documentclass[11pt,fleqn]{article}

\usepackage{a4}
\usepackage{amstext}
\usepackage{amsfonts}
\usepackage{amssymb}
\usepackage{color}
\usepackage{epsfig}
\usepackage{hyperref}

\newcommand{\gtapprox}{\raisebox{-0.5ex}{$\,\stackrel{>}{\scriptstyle\sim}\,$}}
\newcommand{\ltapprox}{\raisebox{-0.5ex}{$\,\stackrel{<}{\scriptstyle\sim}\,$}}

\newcommand{\Oa}{\mathcal{O}(a)}
\newcommand{\bea}{\begin{eqnarray}}
\newcommand{\eea}{\end{eqnarray}}
\newcommand{\beq}{\begin{equation}}
\newcommand{\eeq}{\end{equation}}
\newcommand{\alphah}{\frac{\omega_l}{2}}
\newcommand{\betah}{\frac{\omega_h}{2}}
\def\xlf{\raisebox{+0.2em}{\boldmath{$\chi$}}\hspace{-0.2ex}\raisebox
  {-0.2em}{L}
  \hspace{-1.5ex}\raisebox{+0.14em}{F}\hspace{2mm}}

\begin{document}

% ********************
% ********************
% ********************
% ********************
% ********************

\begin{center}

{\huge \bf Computing $K$ and $D$ meson masses with}

{\huge \bf $N_{\rm f} = 2+1+1$ twisted mass lattice QCD}

\vspace{0.5cm}

SFB/CPP-09-31, LPSC1049, MS-TP-10-10, DESY 10-055, LTH874, LPT-Orsay 10-30, HU-EP-10/19
\vspace{0.5cm}

\textbf{Remi~Baron}$^a$, \textbf{Philippe~Boucaud}$^{b}$, \textbf{Jaume~Carbonell}$^{c}$, \textbf{Vincent~Drach}$^{c}$, \textbf{Federico~Farchioni}$^{d}$, \textbf{Gregorio~Herdoiza}$^e$, \textbf{Karl~Jansen}$^e$, \textbf{Chris~Michael}$^{f}$, \textbf{Istvan~Montvay}$^{g}$, \textbf{Elisabetta~Pallante}$^{h}$, \textbf{Olivier~P\`ene}$^{b}$, \textbf{Siebren~Reker}$^{h}$, \textbf{Carsten~Urbach}$^{i}$, \textbf{Marc~Wagner}$^{j}$, \textbf{Urs~Wenger}$^{k}$

\vspace{0.2cm}

$^a$~CEA, Centre de Saclay, IRFU/Service de Physique Nucl\'eaire, F-91191 Gif-sur-Yvette, France \\
$^b$~Laboratoire de Physique Th\'eorique (B\^at. 210), CNRS et Universit\'e Paris-Sud XI, Centre d'Orsay, 91405 Orsay-Cedex, France \\
$^c$~Laboratoire de Physique Subatomique et Cosmologie, 53 avenue des Martyrs, 38026 Grenoble, France \\
$^d$~Universit\"at M\"unster, Institut f\"ur Theoretische Physik, Wilhelm-Klemm-Stra{\ss}e 9, D-48149 M\"unster, Germany \\
$^e$~NIC, DESY, Platanenallee 6, D-15738 Zeuthen, Germany \\
$^f$~Division of Theoretical Physics, University of Liverpool, L69 3BX Liverpool, United Kingdom \\
$^g$~Deutsches Elektronen-Synchrotron DESY, Notkestr.~85, D-22603 Hamburg, Germany \\
$^h$~Centre for Theoretical Physics, University of Groningen, Nijenborgh 4, 9747 AG Groningen, the Netherlands \\
$^i$~Helmholtz-Institut f{\"u}r Strahlen- und Kernphysik (Theorie) and Bethe Center for Theoretical Physics, Universit{\"a}t Bonn, 53115 Bonn, Germany \\
$^j$~Humboldt-Universit\"at zu Berlin, Institut f\"ur Physik, Newtonstra{\ss}e 15, D-12489 Berlin, Germany \\
$^k$~Albert Einstein Center for Fundamental Physics, Institute for Theoretical Physics, University of Bern, Sidlerstr.\ 5, CH-3012 Bern, Switzerland

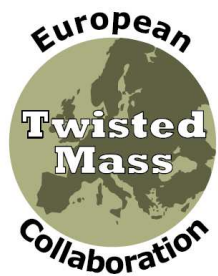
\vspace{0.5cm}

\begin{picture}(0,0)%
\includegraphics{Logo.pstex}%
\end{picture}%
\setlength{\unitlength}{4144sp}%
\begingroup\makeatletter\ifx\SetFigFont\undefined%
\gdef\SetFigFont#1#2#3#4#5{%
  \reset@font\fontsize{#1}{#2pt}%
  \fontfamily{#3}\fontseries{#4}\fontshape{#5}%
  \selectfont}%
\fi\endgroup%
\begin{picture}(1620,1620)(1,-781)
\end{picture}%

\vspace{0.4cm}

\thispagestyle{empty}

May 12, 2010

\end{center}

\vspace{0.1cm}

\newpage

\begin{tabular*}{16cm}{l@{\extracolsep{\fill}}r} \hline \end{tabular*}

\vspace{-0.4cm}
\begin{center} \textbf{Abstract} \end{center}
\vspace{-0.4cm}

We discuss the computation of the mass of the $K$ and $D$ mesons within the framework
of $N_{\rm f} = 2+1+1$ twisted mass lattice QCD from a technical point of view. These quantities
are essential, already at the level of generating gauge configurations, being obvious
candidates to tune the strange and charm quark masses to their physical values. In particular, we
address the problems related to the twisted mass flavor and parity symmetry breaking, which
arise when considering a non-degenerate $(c,\, s)$ doublet. We propose and verify the consistency of three
methods to extract the $K$ and $D$ meson masses in this framework.

\begin{tabular*}{16cm}{l@{\extracolsep{\fill}}r} \hline \end{tabular*}

\thispagestyle{empty}

% ********************
% ********************
% ********************
% ********************
% ********************

\newpage

\setcounter{page}{1}

\section{Introduction}

The framework of maximally twisted mass fermions as an $\mathcal{O}(a)$ improved lattice formulation 
\cite{Frezzotti:2003ni}
has been proved to be highly successful in recent years. The European Twisted Mass Collaboration
(ETMC) has adopted this formulation and has 
carried through a broad research program with $N_{\rm f} = 2$ flavors of mass-degenerate quarks 
in various areas of lattice QCD including 
light meson physics~\cite{Boucaud:2007uk,Boucaud:2008xu,Baron:2009wt}, 
spectroscopy of light baryons~\cite{Alexandrou:2008tn,Alexandrou:2009qu}, 
strange and charm physics~\cite{Blossier:2007vv,Blossier:2009bx,Bertone:2009bu}, 
$B$-physics~\cite{Blossier:2009hg,Blossier:2009gd},
spectroscopy of static-light mesons~\cite{Jansen:2008si,:2010iv}, 
Isgur-Wise functions~\cite{Blossier:2009vy},
meson~\cite{Frezzotti:2008dr,Lubicz:2009ht,DiVita:2009by} and
nucleon~\cite{Alexandrou:2008rp} form factors,
moments of parton distribution functions~\cite{Baron:2007ti},
neutral~\cite{Michael:2007vn} and $\eta'$~\cite{Jansen:2008wv} mesons,
$\omega-\rho$ mass splitting~\cite{McNeile:2009mx},
the vacuum polarization tensor~\cite{Renner:2009by},
pion scattering lengths~\cite{Feng:2009ij},
an investigation of
the $\rho$ meson as a resonance~\cite{Feng:2009ck}
or the non-perturbative renormalization of quark bilinear operators~\cite{Constantinou:2010gr}.

Particular emphasis has been laid on the cut-off effects appearing
at $\mathcal{O}(a^2)$ in the twisted mass formulation at maximal
twist.
%, due to the explicit breaking of both parity and isospin symmetry.  
These effects have been studied theoretically at tree-level of perturbation
theory~\cite{Cichy:2008gk}, and within the Symanzik approach~\cite{Frezzotti:2005gi,Dimopoulos:2009qv}.
These analyses
suggest that isospin breaking effects strongly affect only a limited
set of observables, namely the neutral pion mass and kinematically
related quantities~\cite{Dimopoulos:2009qv}.
The same effects have been numerically investigated 
in the quenched approximation~\cite{Jansen:2005gf,Jansen:2005kk,Abdel-Rehim:2005gz},
with two dynamical
flavors~\cite{Baron:2009wt,Alexandrou:2008tn,Alexandrou:2009qu,Urbach:2007rt,Dimopoulos:2007qy}
and with $N_{\rm f} = 2+1+1$~\cite{Baron:2010LL}.
All numerical results up to date are in agreement with the theoretical conclusions.

%The picture emerging is that for many quantities there is not only
%the expected $\mathcal{O}(a^2)$ scaling towards the continuum limit
%but also that the coefficient multiplying the $\mathcal{O}(a^2)$ term
%is small leading to a mild and well controlled continuum limit
%extrapolation.  The neutral pion plays a special role in this context
%since isospin is broken explicitly in the twisted mass
%formulation. However, the theoretical arguments
%in~\cite{Dimopoulos:2009qv} corroborated by numerical
%results~\cite{Baron:2009wt} suggest that indeed only the neutral pion
%mass is severely affected by these isospin breaking effects.

The studies collected so far suggest that the twisted mass formulation at maximal
twist is a viable realization of QCD on the lattice, with the major advantage of automatic  
$\Oa$ improvement of physical observables, independently of the specific type of operator
considered. Other advantages worth to mention
are that the twisted mass term acts as an infrared regulator of the
theory and that mixing patterns in the renormalisation procedure are
expected to be simplified.
It is hence natural to go one step further and include dynamical strange and 
charm quarks in the simulations. The theoretical
ground for this has been provided in ref.~\cite{Frezzotti:2004wz} and
first feasibility studies have been performed in
ref.~\cite{Chiarappa:2006ae}.  In the last years, we have initiated a
comprehensive research program with dynamical $N_{\rm f}=2+1+1$ flavors of
quarks. Encouraging preliminary results were reported
in~\cite{Baron:2008xa,Baron:2009zq}, while a companion
paper~\cite{Baron:2010LL} presents a more detailed analysis of the light meson sector 
for the ensembles used in this paper. 

%we have demonstrated
%that the charged pion sector and the nucleon mass behave in a very
%similar way as in the $N_{\rm f}=2$ case. In particular, no signs of
%large cut-off effects have been detected for these observables.
%Studies with $N_{\rm f}=2+1+1$ flavors of staggered quarks have been
%presented in~\cite{Bazavov:2009jc,Bazavov:2009wm,Bazavov:2010ru}.

A difficulty arises in $N_{\rm f}=2+1+1$ maximally twisted mass lattice QCD 
when adding a strange and a charm quark, due to the explicit violation of the 
strange and charm flavor quantum number conservation.
%When adding a strange and a charm quark, a good control of the
%heavy-light sector, in particular of the $K$  and $D$ meson masses,
%has to be established. Besides the usual tuning problem to obtain
%physical values of these heavy-light meson masses, a special
%difficulty arises in $N_{\rm f}=2+1+1$ maximally twisted mass lattice
%QCD due to the explicit violation of the strange and charm flavor
%quantum number conservation occurring in this formulation.
%%%Istvan
At any non-vanishing value of the lattice spacing, the latter leads to the
contamination of correlators by unphysical contributions from intermediate
states carrying the wrong quantum numbers.
%%%Istvan
%At any non-vanishing value of the lattice spacing, these effects lead to the appearance 
%of unphysical states which contaminate the correlators one is interested in; \\
%\textbf{XXXXX Marc: I think this is not correct ... states qualitatively remain the same ... 
%there is only a smaller number of well-defined quantum numbers. XXXXX} \\
Moreover, transitions that are not allowed in continuum QCD become possible,
the consequence being that stable states in the continuum with respect to strong interactions, 
such as the $D$ meson, become resonances.

In this paper, we provide algorithmic and methodological tools to
tackle the problem. In particular, we present three
techniques, a generalized eigenvalue problem, multiple exponential
fits, and enforcing parity and flavor symmetry restoration, to compute
the physical $K$ and $D$ meson masses. As we will demonstrate below,
we find that with all three methods these masses can be extracted and
results agree among the three methods.  
The paper is conceived as a technical report on these methods, which can
 in general be applied whenever flavor symmetry breaking occurs.
Efforts to implement these techniques in combination with a flavor diagonal
 Osterwalder-Seiler valence quark action, see  
e.g.~\cite{Blossier:2007vv,Frezzotti:2004wz,Pena:2004gb,AbdelRehim:2006ve}, 
are ongoing.

The paper is organized as follows. In section \ref{sec:setup} we define the setup, the operators used, and the optimization of the correlation matrices.  
Section \ref{SEC001} describes the determination of the $K$ and $D$ meson masses with the three methods. We conclude in 
section \ref{sec:concl}. 

% ********************
% ********************
% ********************
% ********************
% ********************

\section{Simulation setup}
\label{sec:setup}
% **********

\subsection{$N_{\rm f} = 2+1+1$ twisted mass lattice QCD}

This work is based on sets of configurations generated by the ETM collaboration~\cite{Baron:2008xa,Baron:2009zq} with the
Iwasaki gauge action~\cite{Iwasaki:1985we} and $N_{\rm f} = 2+1+1$ flavors of twisted mass quarks. The light degenerate $(u,d)$ quark doublet is described by the standard twisted mass action~\cite{Frezzotti:2000nk}
\begin{eqnarray}\label{eq:act_l}
S_{\mathrm{F},\textrm{light}}[\chi^{(l)},\bar{\chi}^{(l)},U] \ \ = \ \ a^4 \sum_x \bar{\chi}^{(l)}(x) \Big(D_\textrm{W}(m_0) + i \mu \gamma_5 \tau_3\Big) \chi^{(l)}(x) ,
\end{eqnarray}
while for the $(c,s)$ doublet the twisted mass formulation for non-degenerate quarks of~\cite{Frezzotti:2003xj} has been used:
\begin{eqnarray}
\label{EQN345} S_{\mathrm{F},\textrm{heavy}}[\chi^{(h)},\bar{\chi}^{(h)},U] \ \ = \ a^4 \sum_x \bar{\chi}^{(h)}(x) \Big(D_\textrm{W}(m_0) + i \mu_\sigma \gamma_5 \tau_1 + \tau_3 \mu_\delta\Big) \chi^{(h)}(x) .
\end{eqnarray}
In both cases $D_\mathrm{W}$ denotes the standard Wilson Dirac operator
\begin{eqnarray}\label{eq:lat_der}
D_\mathrm{W}(m_0) \ \ = \ \ \frac{1}{2} \Big(\gamma_\mu \Big(\nabla_\mu + \nabla^\ast_\mu\Big) - a \nabla^\ast_\mu \nabla_\mu\Big) + m_0 ,
\end{eqnarray}
while $\chi^{(l)} = (\chi^{(u)},\chi^{(d)})$ and $\chi^{(h)} = (\chi^{(c)},\chi^{(s)})$ are the quark fields in the so-called twisted basis. For reasons explained in~\cite{Chiarappa:2006ae} the same value of the standard quark mass parameter $m_0$ has been used in both sectors.

When tuning the theory to maximal twist, automatic $\mathcal{O}(a)$ improvement for physical quantities applies~\cite{Frezzotti:2003xj,Frezzotti:2003ni}. This tuning has been done by adjusting $m_0$ such that the PCAC quark mass in the light quark sector vanishes~\cite{Baron:2010LL},
\begin{eqnarray}
\label{eq:PCAC} a m_{\chi^{(l)}}^\textrm{PCAC} \ \ = \ \ \frac{\sum_{\mathbf{x}} \Big\langle \partial^\ast_0 {A^{(l) +}_0}(x) P^{(l) -}(y) \Big\rangle}{2 \sum_{\mathbf{x}} \Big\langle P^{(l) +}(x) P^{(l) -}(y) \Big\rangle} \ \ = \ \ 0\, ,
\end{eqnarray}
with the bilinears defined as
\begin{eqnarray}
\label{eq:PCAC_} A^{(l) +}_\mu \ \ = \ \ \bar\chi^{(u)} \gamma_\mu \gamma_5 \chi^{(d)} \quad , \quad P^{(l) +} \ \ = \ \ \bar\chi^{(u)} \gamma_5 \chi^{(d)} \quad , \quad P^{(l) -} \ \ = \ \ \bar\chi^{(d)} \gamma_5 \chi^{(u)} .
\end{eqnarray}

%The heavy doublet mass parameters $\mu_\sigma$ and $\mu_\delta$ have been chosen 
%in our simulations such that the $s$ and $c$ quark masses are in vicinity of the respective physical values. 
At maximal twist, in a massless quark renormalization scheme, the renormalized quark masses are related to the bare parameters $\mu_\sigma$ and $\mu_\delta$ by~\cite{Frezzotti:2003xj}
\begin{eqnarray}
\label{eq:csmasses} m^R_{s} \ \ = \ \ Z_P^{-1} \bigg(\mu_\sigma -
\frac{Z_P}{Z_S} \mu_\delta\bigg) \qquad , \qquad
m^R_{c} \ \ = \ \ Z_P^{-1} \bigg(\mu_\sigma + \frac{Z_P}{Z_S}\, ,
\mu_\delta\bigg)
\end{eqnarray}
where $Z_P$ and $Z_S$ are the renormalization constants of the non-singlet pseudoscalar and scalar densities in a massless quark scheme, namely for $N_{\rm f} = 4$ massless Wilson lattice QCD.

The values of $\mu_\sigma$ and $\mu_\delta$ have been adjusted in our simulations 
by requiring that the simulated kaon and D meson mass approximately assume their physical values
\cite{Baron:2010LL}.
% Alternatively, through the knowledge of the
% ratio of the renormalization factors $Z_P/Z_S$, the ratio of the
% renormalized charm and strange quark masses can be set to satisfy
% $(m_c/m_s)_R \approx 10$ according to eq.~(\ref{eq:csmasses}). An
% estimate of the $Z_P/Z_S$ ratio has been
% obtained by a method explained in detail in section~\ref{SEC422},
% while dedicated runs to perform the non-perturbative renormalization
% are in progress.
% {Fed: \bf I would remove this, since we will never use this method for tuning the masses in the sea, 
% but only for providing an input for OS, which is beyond what we want to exaplain here}
For this study we consider two ensembles, one from each of the
currently simulated $\beta$ values, $\beta=1.90$ and
$\beta=1.95$~\cite{Baron:2008xa,Baron:2009zq,Baron:2010LL},
with a light pseudoscalar mass $m_\textrm{PS} \approx 320 \, 
\textrm{MeV}$ in both cases, see Table~\ref{TAB001}.

% pion masses in physical units

% > mPS = 0.14142
% > a = 0.086
% > mPS / a * 197
% [1] 323.9505

% > mPS = 0.12602
% > a = 0.078
% > mPS / a * 197
% [1] 318.2813

\begin{table}[htb]
\begin{center}

\begin{tabular}{|c|c|c|c|c|c|c|c|c|c|}
\hline
 & & & & & & & & & \vspace{-0.40cm} \\
 & & & & & & & $a$ & $m_\textrm{PS}$ & \# of \\
Ensemble & $\beta$ & $(L/a)^3 \times T/a$ & $a\mu$ & $\kappa$ & $a\mu_\sigma$ & $a\mu_\delta$ & in $\textrm{fm}$ 
& in $\textrm{MeV}$ & gauges \\
 & & & & & & & & & \vspace{-0.40cm} \\
\hline
 & & & & & & & & & \vspace{-0.40cm} \\
A40.32 & $1.90$ & $32^3 \times 64$ & $0.0040$ & $0.163270$ & $0.150$ & $0.190$ & $0.086$ & $324$ & $1003$ \\
B35.32 & $1.95$ & $32^3 \times 64$ & $0.0035$ & $0.161240$ & $0.135$ & $0.170$ & $0.078$ & $318$ & $1042$\vspace{-0.40cm} \\
 & & & & & & & & & \\
\hline
\end{tabular}

\caption{\label{TAB001}Summary of the ensembles considered in this paper, more details in~\cite{Baron:2008xa,Baron:2009zq,Baron:2010LL}.}

\end{center}
\end{table}

% **********

\subsection{Meson creation operators and trial states}

% *****

\subsubsection{\label{sec:d_f_structure}Quantum numbers, physical basis and twisted basis}

% K^+ = \bar{s} u   <-->   K^- = \bar{u} s
% K^0 = \bar{s} d   <-->   \bar{K}^0 = bar{d} s
% 
% D^+ = \bar{d} c   <-->   D^- = \bar{c} d
% D^0 = \bar{u} c   <-->   \bar{D}^0 = bar{c} u

We are concerned with computing the mass of the $K$ meson, $m_K$, and of the $D$ meson, 
$m_D$, within the setup defined by eqs.\ (\ref{eq:act_l}) to (\ref{eq:lat_der}). 
Both mesons have total angular momentum $J = 0$ and parity $\mathcal{P = -}$. Their quark content is e.g.\ $\bar{K}^0 \equiv \bar{d} s$ and $D^+ \equiv \bar{d} c$.

Neither heavy flavor nor parity are exact
symmetries in $N_{\rm f} = 2+1+1$ twisted mass lattice QCD at finite
lattice spacing. In particular, the $\tau_1$-coupling term
in eq.~(\ref{EQN345}) violates the conservation of the strange and charm
flavor quantum numbers. Consequently, instead of four different
heavy-light meson sectors $(s,-)$, $(s,+)$, $(c,-)$ and $(c,+)$ there
is only a single mixed flavor-parity sector $(s/c,-/+)$. Problems
arise in particular when one tries to determine $m_D$. In continuum QCD the $D$
meson is the lowest state in the $(c,-)$ sector, while in our setup it
is a highly excited state in the combined sector $(s/c,-/+)$. Notice that, besides
the $K$ meson, there are a radially excited $K$ state ($K(1460)$),
possibly strange mesons with positive parity ($K_0^\ast(800)$,
$K_0^\ast(1430)$) and a number of multi particle states $K/K_0^\ast +
n \times \pi$~\cite{PDG}. Hence, for a clean extraction of $m_D$ one has to
consider sufficiently large correlation matrices, which are able to
resolve all these low lying states. 
This is possible in principle. In practice, the separation of 
the excited states would require the determination of correlation
matrices with extremely high statistical precision. At our currently
available statistics, this route seems not to be viable.

Our approach is instead based on the observation that parity and heavy flavor symmetries are restored in the continuum limit, where the twisted mass theory is expected to reproduce QCD with $N_{\rm f} = 2+1+1$ quark flavors. In this limit, operators with definite parity \cite{Frezzotti:2003ni} and flavor quantum numbers projecting onto the physical meson states can be reconstructed (cf.\ section~\ref{SEC422}). As it is shown in the following, these operators can be defined as linear combinations of bilinears of the lattice quark fields in the twisted basis.

In the continuum, or in any chirality preserving lattice formulation~\cite{Frezzotti:2000nk}, 
the twist transformation relating the physical quark fields $\psi$ and the twisted quark fields $\chi$ reads
\begin{eqnarray}
\label{eq:twist} & & \hspace{-0.7cm} \psi^{(l)} \ \ = \ \ e^{i \omega_l \gamma_5 \tau_3 / 2} \chi^{(l)} \quad , \quad \bar\psi^{(l)} \ \ = \ \ \bar\chi^{(l)} e^{i \omega_l \gamma_5 \tau_3 / 2} \\
\label{eq:twist_} & & \hspace{-0.7cm} \psi^{(h)} \ \ = \ \ e^{i \omega_h \gamma_5 \tau_1 / 2} \chi^{(h)} \quad , \quad \bar\psi^{(h)} \ \ = \ \ \bar\chi^{(h)} e^{i \omega_h \gamma_5 \tau_1 / 2} ,
\end{eqnarray}
where $\omega_{l,h}$ are the twist angles in the light and heavy quark sector, respectively.
Analogous relations hold for operators projecting, in the continuum limit, on trial states with definite heavy flavor and parity quantum numbers. In the physical basis, such operators can  be chosen according to\footnote{For definiteness we 
identify the light flavor with $d$.}
\begin{eqnarray}
\label{EQN001_ph} & & \hspace{-0.7cm} \mathcal{O}_\textrm{ph} \ \ = \ \ \left( \! \! \begin{array}{c}
  \mathcal{O}_\textrm{ph}^{(s,\gamma_5)} \\
  \mathcal{O}_\textrm{ph}^{(c,\gamma_5)} \\
  \mathcal{O}_\textrm{ph}^{(s,1)} \\
  \mathcal{O}_\textrm{ph}^{(c,1)}
\end{array} \! \! \right ) \ \ = \ \ \left( \! \! \begin{array}{c}
  +i \bar{\psi}^{(d)} \gamma_5 \psi^{(s)} \\
  -i \bar{\psi}^{(d)} \gamma_5 \psi^{(c)} \\
  +\bar{\psi}^{(d)} \psi^{(s)} \\
  -\bar{\psi}^{(d)} \psi^{(c)}
\end{array} \! \! \right )
\end{eqnarray}
The twist rotations in eqs.~(\ref{eq:twist}) and (\ref{eq:twist_}) relate the twisted basis operators
\begin{eqnarray}
\label{EQN001} & & \hspace{-0.7cm} \mathcal{O}_\chi \ \ = \ \ \left( \! \! \begin{array}{c}
  \mathcal{O}_\chi^{(s,\gamma_5)} \\
  \mathcal{O}_\chi^{(c,\gamma_5)} \\
  \mathcal{O}_\chi^{(s,1)} \\
  \mathcal{O}_\chi^{(c,1)}
\end{array} \! \! \right ) \ \ = \ \ \left( \! \! \begin{array}{c}
  +i \bar{\chi}^{(d)} \gamma_5 \chi^{(s)} \\
  -i \bar{\chi}^{(d)} \gamma_5 \chi^{(c)} \\
  +\bar{\chi}^{(d)} \chi^{(s)} \\
  -\bar{\chi}^{(d)} \chi^{(c)}
\end{array} \! \! \right ) 
\end{eqnarray}
to the physical operators of eq.~(\ref{EQN001_ph}) as follows
\begin{eqnarray}
\label{eq:op_phys} \mathcal{O}_\textrm{ph} \ \ = \ \ \mathcal{M}(\omega_l,\omega_h)\, \mathcal{O}_\chi \quad , \quad \mathcal{O}_\textrm{ph}^\dagger \ \ = \ \ \mathcal{O}_\chi^\dagger\, \mathcal{M}^T(\omega_l,\omega_h)\, ,
\end{eqnarray}
with the orthogonal twist rotation matrix given by
\begin{eqnarray}
\label{eq:mix} \mathcal{M}(\omega_l,\omega_h) \ \ = \ \ \left(\begin{array}{cccc}
\phantom{-} \cos\alphah \cos\betah & - \sin\alphah \sin\betah & - \sin\alphah \cos\betah & - \cos\alphah \sin\betah \\ [1mm]
- \sin\alphah \sin\betah & \phantom{-} \cos\alphah \cos\betah & - \cos\alphah \sin\betah & - \sin\alphah \cos\betah \\ [1mm]
\phantom{-} \sin\alphah \cos\betah & \phantom{-} \cos\alphah \sin\betah & \phantom{-} \cos\alphah \cos\betah & - \sin\alphah \sin\betah \\ [1mm]
\phantom{-} \cos\alphah \sin\betah & \phantom{-} \sin\alphah \cos\betah & - \sin\alphah \sin\betah & \phantom{-} \cos\alphah \cos\betah
\end{array}\right)\, .
\end{eqnarray}
However, when using the Wilson lattice formulation, the operators in eq.~(\ref{EQN001}), with and without a $\gamma_5$ matrix, renormalize differently due to the explicit breaking of chiral symmetry. This implies that, to be able to build a representation of the chiral group, renormalization factors must explicitly be taken into account, and eq.~(\ref{eq:op_phys}) only holds for the renormalized counterparts
\begin{eqnarray}
\label{eq:op_phys_ren} 
\mathcal{O}_\textrm{ph}^R \ \ = \ \ \mathcal{M}(\omega_l,\omega_h)\, 
\mathcal{O}_\chi^R \quad , \quad (\mathcal{O}_\textrm{ph}^R)^\dagger \ \ 
= \ \ (\mathcal{O}_\chi^R)^\dagger\, \mathcal{M}^T(\omega_l,\omega_h) ,
\end{eqnarray}
where the bilinears in eq.~(\ref{EQN001}) have been replaced by their renormalized versions,
\begin{eqnarray}
\label{eq:vec_chi_ren} \mathcal{O}_\chi^R \ \ = \ \ \textrm{diag}\Big(Z_P,Z_P,Z_S,Z_S\Big)\,\mathcal{O}_\chi \ \ = \ \ \left( \! \! \begin{array}{c}
  Z_P\, \mathcal{O}_\chi^{(s,\gamma_5)} \\
  Z_P\, \mathcal{O}_\chi^{(c,\gamma_5)} \\
  Z_S\, \mathcal{O}_\chi^{(s,1)} \\
  Z_S\, \mathcal{O}_\chi^{(c,1)}
\end{array} \! \! \right ) \, ,
\end{eqnarray}
and $Z_P$ and $Z_S$ are the same renormalization factors as in (\ref{eq:csmasses}).
At maximal twist, i.e.\ $\omega_l=\omega_h=\pi/2$, one has 
\begin{eqnarray}
\label{eq:op_phys_ren_mt} \left( \! \! \begin{array}{c}
  \mathcal{O}_\textrm{ph}^{(s,\gamma_5)} \\
  \mathcal{O}_\textrm{ph}^{(c,\gamma_5)} \\
  \mathcal{O}_\textrm{ph}^{(s,1)} \\
  \mathcal{O}_\textrm{ph}^{(c,1)}
\end{array} \! \! \right)^{\!\!\!R} \ \ = \ \ \frac{1}{2} \left(\begin{array}{rrrr}
  1 & -1 & -1 & -1 \\ 
 -1 &  1 & -1 & -1 \\ 
  1 &  1 &  1 & -1 \\ 
  1 &  1 & -1 &  1
\end{array}\right) \left( \! \! \begin{array}{c}
  Z_P\, \mathcal{O}_\chi^{(s,\gamma_5)} \\
  Z_P\, \mathcal{O}_\chi^{(c,\gamma_5)} \\
  Z_S\, \mathcal{O}_\chi^{(s,1)} \\
  Z_S\, \mathcal{O}_\chi^{(c,1)}
\end{array} \! \! \right ) \, .
\end{eqnarray}
A third definition of the quark fields will be useful in the following
(where maximal twist applies),
obtained by rotating the lattice $\chi$-fields via eqs.~(\ref{eq:twist}) and (\ref{eq:twist_}), where now 
$\omega_l=\omega_h=\pi/2$.
The rotated fields would reproduce the physical ones in a theory with
exact chiral symmetry and $Z_P=Z_S$.
In the present formulation with broken chiral symmetry,
they define instead a ``pseudo physical basis'' (ppb). We denote the rotated fields with
$\psi_\textrm{ppb}^{(l,\,h)}$ and introduce 
the operator bilinears in this basis
\begin{eqnarray}
\label{EQN001_ppb} & & 
\hspace{-0.7cm} \mathcal{O}_\textrm{ppb} \  =  \ \left( \! \! \begin{array}{c}
  \mathcal{O}_\textrm{ppb}^{(s,\gamma_5)} \\
  \mathcal{O}_\textrm{ppb}^{(c,\gamma_5)} \\
  \mathcal{O}_\textrm{ppb}^{(s,1)} \\
  \mathcal{O}_\textrm{ppb}^{(c,1)}
\end{array} \! \! \right ) \ =  \ \left( \! \! \begin{array}{c}
  +i \bar{\psi}_\textrm{ppb}^{(d)} \gamma_5 \psi_\textrm{ppb}^{(s)} \\
  -i \bar{\psi}_\textrm{ppb}^{(d)} \gamma_5 \psi_\textrm{ppb}^{(c)} \\
  +\bar{\psi}_\textrm{ppb}^{(d)} \psi_\textrm{ppb}^{(s)} \\
  -\bar{\psi}_\textrm{ppb}^{(d)} \psi_\textrm{ppb}^{(c)}
\end{array} \! \! \right )
\ \ = \ \ \frac{1}{2} \left(\begin{array}{rrrr}
  1 & -1 & -1 & -1 \\ 
 -1 &  1 & -1 & -1 \\ 
  1 &  1 &  1 & -1 \\ 
  1 &  1 & -1 &  1
\end{array}\right) \left( \! \! \begin{array}{c}
  \mathcal{O}_\chi^{(s,\gamma_5)} \\
  \mathcal{O}_\chi^{(c,\gamma_5)} \\
  \mathcal{O}_\chi^{(s,1)} \\
  \mathcal{O}_\chi^{(c,1)}
\end{array} \! \! \right )\, ,
\end{eqnarray}
otherwise written as
\begin{eqnarray}
\label{eq:op_ppb} 
\mathcal{O}_\textrm{ppb} \ \ = \ \ \mathcal{M}(\pi /2,\pi /2)\, 
\mathcal{O}_\chi \ \ \equiv \ \ \mathcal{M}_{\textrm{mt}}\, \mathcal{O}_\chi\, .
\end{eqnarray}
The physical operators defined in eq.~(\ref{eq:op_phys_ren}), and eq.~(\ref{eq:op_phys_ren_mt}) at maximal twist, project 
onto states that converge to states with definite flavor and parity quantum numbers in 
the continuum limit. 
Since we aim to determine the ground states of the physical system, in at least two of 
the four sectors, it is appropriate to first build the correlation matrices in terms of 
the building blocks given in eq.~(\ref{EQN001}).
We also project to zero momentum by summing over all lattice sites at fixed Euclidean 
time $t$,
\begin{eqnarray}
\label{EQN001_old} \mathcal{O}_\chi^{(h,\Gamma)}(t) \ \ = \ \ \eta_\Gamma\,\sum_\mathbf{x} \bar{\chi}^{(d)}(\mathbf{x},t) \Gamma \chi^{(h)}(\mathbf{x},t) \  , \  h \in \{ s \, , \, c \} \ , \  \Gamma \in \{ \gamma_5 \, , \, 1 \} \ \ (\eta_1=\pm 1,\eta_{\gamma_5}=\pm i)\ .
\end{eqnarray}
The corresponding trial states
\begin{eqnarray}
\label{EQN101} | \phi_\chi^{(h,\Gamma)}(t) \rangle \ \ = \ \ \Big(\mathcal{O}_\chi^{(h,\Gamma)}(t)\Big)^\dagger | \Omega \rangle
\end{eqnarray}
enter the correlation matrices
\begin{eqnarray}
\label{EQN003} C_{(h_2,\Gamma_2),(h_1\Gamma_1)}(t_2-t_1) =  \langle \phi_\chi^{(h_2,\Gamma_2)}(t_2) | \phi_\chi^{(h_1,\Gamma_1)}(t_1) \rangle 
=   \langle\Omega| \Big(\mathcal{O}_\chi^{(h_2,\Gamma_2)}(t_2)\Big) \Big(\mathcal{O}_\chi^{(h_1,\Gamma_1)}(t_1)\Big)^\dagger|\Omega\rangle\  ,
\end{eqnarray}
and we introduce the shorthand matrix notation for later use 
\begin{eqnarray}
\label{eq:corr_mat} C(t_2-t_1) \ \ = \ \ \Big\langle \mathcal{O}(t_2) \otimes (\mathcal{O}(t_1))^\dagger \Big\rangle \ .
\end{eqnarray}
Notice also that, due to the discrete symmetries of the twisted mass action
in eqs.~(\ref{eq:act_l}) and (\ref{EQN345}), the correlation $C(t_2-t_1)$ 
is a real and symmetric matrix.
Eqs. (\ref{EQN001_old}) to (\ref{eq:corr_mat}) can also be generalized to the case of more
operators, as for example operators with different levels of smearing (see the next section)
or Dirac structure. 
In this case $C(t_2-t_1)$ will be a $D\times D$ matrix ($D=4\times n$)
defined by the larger operator set. An application of this kind will be considered in 
section~\ref{sec:method_2}.
One can easily obtain another set of independent meson creation
operators with identical quantum numbers by replacing $\Gamma
\rightarrow \gamma_0 \Gamma$. We found, however, that the
corresponding trial states have worse overlaps to the low lying states
of interest. Therefore, we do not consider these operators in the
following.
To improve the signal-to-noise ratio, we have computed the correlators in 
eq.~(\ref{EQN003}) by using the one-end trick~\cite{Boucaud:2007uk,Boucaud:2008xu}.
%Starting from the correlation matrices (\ref{EQN003}) we extract the meson masses 
%$m_K$ and $m_D$ by applying three different methods, 
%which will be explained in section \ref{SEC001}.

% *****

\subsubsection{Operator optimization by means of smearing}

To optimize the overlap of the trial states in eq.~(\ref{EQN101}) with the physical 
$K$ and $D$ mesons, we resort to standard smearing techniques. 
We use Gaussian smeared quark fields, with APE smeared spatial links. 
Additional details can be found in \cite{Jansen:2008si}, 
where the same setup has been used.

\begin{figure}[htb]
\begin{center}
\input{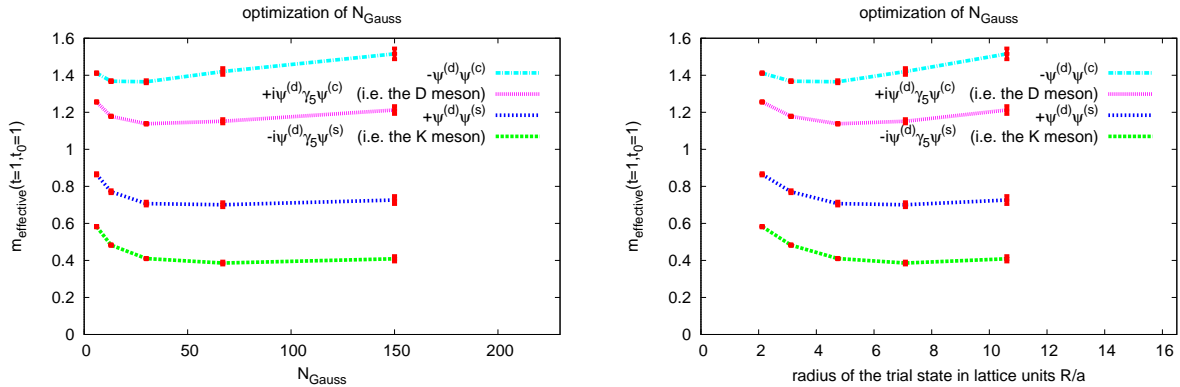}
\caption{\label{FIG001}
\textbf{a)}~The effective masses $m_\textrm{effective}^{(n)}(t=1,t_0=1)$ (cf.\ eq.\ (\ref{EQN005})) for the trial states defined in eq.\ (\ref{EQN101}) as functions of $N_\textrm{Gauss}$ for $\beta = 3.90$, $L^3 \times T = 24^3 \times 48$ and $\mu = 0.0040$ with $\kappa_\textrm{Gauss} = 0.5$, $N_\textrm{APE} = 10$, $\alpha_\textrm{APE} = 0.5$.
\textbf{b)}~The same effective masses as a function of the radius of the trial states in lattice units $R/a$ with $\kappa_\textrm{Gauss} = 0.5$.
}
\end{center}
\end{figure}

We have optimized the smearing by computing effective masses at $t = 1$
and $t_0 = 1$ (cf.\ (\ref{EQN005})), where excited states are
suppressed the least, for different values of $N_\textrm{Gauss}$, and 
$\kappa_\textrm{Gauss} = 0.5$, $N_\textrm{APE} = 10$,
$\alpha_\textrm{APE} = 0.5$ kept fixed. This optimization is essentially independent on the lattice volume and on the light quark mass. Results for $\beta = 3.90$, $L^3 \times T = 24^3 \times 48$ and $\mu = 0.0040$ are reported in Figure~\ref{FIG001}. Although the suppression of excited states only weakly depends
on $N_\textrm{Gauss}$ and, therefore, on the width of the
corresponding trial states, it is obvious that the $D$ meson has a
somewhat smaller width than the $K$ meson. Since the $D$ meson is
heavier and hence more difficult to compute, we focus on optimizing
the overlap with the $D$ meson state and choose $N_\textrm{Gauss} = 30$. An estimate
of the corresponding trial state radius $R$ can be obtained via~\cite{Jansen:2008si}
\begin{eqnarray}
\frac{R}{a} \ \ = \ \ \bigg(\frac{N_\textrm{Gauss} \kappa_\textrm{Gauss}}{1 + 6 \kappa_\textrm{Gauss}}\bigg)^{1/2}\,,
\end{eqnarray}
yielding $R_K \approx 7 a \approx 0.60 \, \textrm{fm}$ and $R_D \approx 5 a \approx 0.43 \, \textrm{fm}$ (cf.\ Figure~\ref{FIG001}b).
\begin{figure}[htb]
\begin{center}
\input{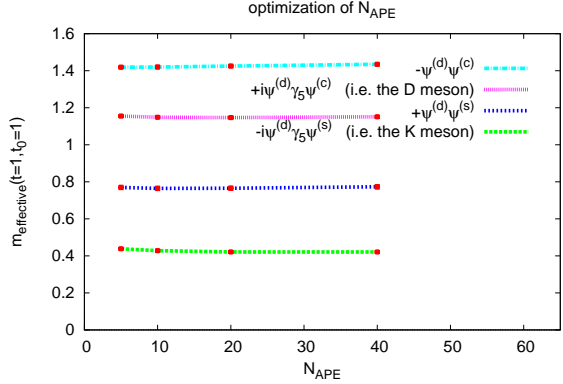}
\caption{\label{FIG002} The effective masses $m_\textrm{effective}^{(n)}(t=1,t_0=1)$ as 
in Figure~\ref{FIG001}, 
as a function of $N_\textrm{APE}$ for $\beta = 3.90$, $L^3 \times T = 24^3 \times 48$ and $\mu = 0.0100$ with $\alpha_\textrm{APE} = 0.5$, 
$N_\textrm{Gauss} = 30$, $\kappa_\textrm{Gauss} = 0.5$.}
\end{center}
\end{figure}
A similar optimization for the parameter $N_\textrm{APE}$ shows essentially no dependence on the ground state overlap. This is exemplified in Figure~\ref{FIG002} corresponding to $\beta = 3.90$, $L^3 \times T = 24^3 \times 48$ and $\mu = 0.0100$.

We end up with the following optimized set of smearing parameters for ensemble A40.32:
\begin{eqnarray}
\label{EQN002} N_\textrm{Gauss} = 30 \quad , \quad \kappa_\textrm{Gauss} = 0.5 
\quad , \quad N_\textrm{APE} = 10 \quad , \quad \alpha_\textrm{APE} = 0.5 .
\end{eqnarray}
Given the rather mild dependence of the ground state overlap on $N_\textrm{Gauss}$ and 
$N_\textrm{APE}$, we use the set of 
 parameters in eq.~(\ref{EQN002}) also for the ensemble B35.32, with only slightly 
different lattice spacing.
Sometimes in the following of this paper, we will also consider 
correlation matrices made of local operators, 
or  mixed local and smeared operators. However, the final determination of all masses will 
exclusively be obtained 
with the correlation matrix made of the smeared operators, with the optimized 
smearing parameters of eq.~(\ref{EQN002}).

% ********************
% ********************
% ********************
% ********************
% ********************

\section{\label{SEC001}Computation of $m_K$ and $m_D$}

%As explained in section~\ref{sec:d_f_structure}, heavy flavor as well as parity are not 
%good quantum numbers of twisted mass lattice QCD as defined 
%by Eqs. (\ref{eq:act_l}) to (\ref{eq:lat_der}). 
In contrast to parity and flavor 
conserving lattice formulations, as the standard Wilson lattice QCD, it is not possible 
to compute correlation functions restricted to a single parity and heavy flavor sector 
in our $N_{\rm f} = 2+1+1$ twisted mass framework, as outlined in 
section~\ref{sec:d_f_structure}.
While the determination of $m_K$ is anyway straightforward, since the kaon is the 
lowest state in the combined heavy flavor and parity sector, the extraction of 
$m_D$ remains rather problematic, being the $D$ meson a highly excited state. 
Besides computing $m_K$ with high precision, we attempt in the following to estimate 
$m_D$ without computing the full low-lying spectrum. We present and compare three different 
methods, all based on the fact that both heavy flavor symmetry and parity are only 
weakly broken, by terms of $\mathcal{O}(a)$. The three methods yield a consistent picture.

% **********

\subsection{\label{SEC003}Method~1: solving a generalized eigenvalue problem}

We consider $4 \times 4$ correlation matrices, as defined in eq.~(\ref{EQN003}),
 computed with the twisted basis operators of eq.~(\ref{EQN001}) and the 
optimized smearing parameters given in eq.~(\ref{EQN002}). We then solve the 
generalized eigenvalue problem
\begin{eqnarray}
\label{EQN004} \sum_k C_{j k}(t) v_k^{(n)}(t,t_0) \ \ = \ \ \sum_k C_{j k}(t_0) v_k^{(n)}(t,t_0) \lambda^{(n)}(t,t_0) \quad , \quad t \ \ \equiv \ \ t_2-t_1
\end{eqnarray}
where $k$ runs over the set $(h,\Gamma ),\, h=c,s,\, \Gamma =\pm$, 
and obtain the four effective masses $m_\textrm{effective}^{(n)}$, with $n=0,\ldots ,3$,
 by solving~\cite{Blossier:2009kd}
\begin{eqnarray}
\label{EQN005} \frac{\lambda^{(n)}(t,t_0)}{\lambda^{(n)}(t+1,t_0)} \ \ = \ \ \frac{e^{-m_\textrm{effective}^{(n)}(t,t_0) t} + e^{-m_\textrm{effective}^{(n)}(t,t_0) (T-t)}}{e^{-m_\textrm{effective}^{(n)}(t,t_0) (t+1)} + e^{-m_\textrm{effective}^{(n)}(t,t_0) (T-(t+1))}}\, ,
\end{eqnarray}
with $T$ the temporal extension of the periodic lattice.

To interpret these effective masses, we assume that heavy flavor and parity breaking effects are small. Indeed they are only $\mathcal{O}(a)$, since they originate from the flavor non-diagonal and parity odd Wilson term, which is proportional to the lattice spacing. Consequently, for vanishing lattice spacing, where heavy flavor and parity are exact symmetries, these correlation matrices would be diagonal in the physical basis, because the operators in eq.\ (\ref{EQN001_ph}) would excite orthogonal trial states. Thus, solving the generalized eigenvalue 
problem as stated in eq.~(\ref{EQN004}) would directly provide the four effective masses 
with definite heavy flavor and parity.
%be straightforward and each of the four effective masses (\ref{EQN005}) would correspond 
%to a different heavy flavor/parity sector. 
In particular, one of them would have associated quantum numbers $(c,\,-)$ 
and would approach a plateau for large temporal separation to be identified 
with the $D$ meson mass.

At finite lattice spacing in the presence of heavy flavor and parity breaking
the four effective masses will approach the masses of the four lowest states in the 
mixed sector $(s/c,-/+)$ for large temporal separations.
The $D$ meson is not among those states: $K$ and $K_0^\ast$, the radial excitations 
and $K/K_0^\ast + n \times \pi$ states are lighter than the $D$. 
At intermediate times, however, one of the four effective masses should still be dominated by the $D$ meson and the corresponding plateau will give a measure of
 $m_D$.

To identify the heavy flavor and parity content of the four 
effective masses, we first note that the trial state corresponding to the $n$-th effective mass is
\begin{eqnarray}\label{EQN010}
 | \phi_\chi^{(n)}(t) \rangle \ \ = \ \ \sum_k\,v_k^{(n)}(t,t_0)\, 
\Big(\mathcal{O}_\chi^{(k)}(t)\Big)^\dagger|\Omega\rangle\ , 
\end{eqnarray}
%
%%where $k$ is a collective index for $(h,\Gamma)$, see (\ref{EQN001_old}).
When the relations $\omega_l = \omega_h = \pi / 2$ and $Z_P / Z_S =1$ are approximately 
fulfilled, one can rotate to the pseudo physical basis. By
inserting eq.\ (\ref{EQN001_ppb}) into the trial state in (\ref{EQN010})
and using the orthogonality of the twist rotation matrix $\mathcal{M}_{\textrm{mt}}$ at 
maximal twist of eq.~(\ref{eq:op_ppb}), yields 
\begin{eqnarray}\label{eq:exp_ppb}
| \phi_\chi^{(n)}(t) \rangle \ \ = \ \ \sum_k\,
\Big(\mathcal{M}_{\textrm{mt}}\,v^{(n)}(t,t_0)\Big)_k
\Big(\mathcal{O}^{(k)}_\textrm{ppb}(t)\Big)^\dagger|\Omega\rangle\, .
\end{eqnarray}
By sorting the terms in eq.~(\ref{eq:exp_ppb}) according to the pseudo physical basis 
 states $(\mathcal{O}^{(k)}_\textrm{ppb})^\dagger|\Omega\rangle$,
the approximate heavy flavor and parity contents of the trial state corresponding to the 
$n$-th effective mass can be read off, and it is given by  
$c_{(h,\Gamma)}^{(n)}\propto|(\mathcal{M}_{\textrm{mt}}\,v^{(n)}(t,t_0))_{(h,\Gamma)}|^2$.
Explicitly,
\begin{eqnarray}
\label{EQN644} & & \hspace{-0.7cm} c_{(s,\gamma_5)}^{(n)}
\ \ = \ \ \frac{1}{N} \,\Big|
  + v_{(s,\gamma_5)}^{(n)}
  - v_{(c,\gamma_5)}^{(n)}
  - v_{(s,1)}^{(n)}
  - v_{(c,1)}^{(n)}
\Big|^2 \\
& & \hspace{-0.7cm} c_{(c,\gamma_5)}^{(n)}
\ \ = \ \ \frac{1}{N} \,\Big|
  - v_{(s,\gamma_5)}^{(n)}
  + v_{(c,\gamma_5)}^{(n)}
  - v_{(s,1)}^{(n)}
  - v_{(c,1)}^{(n)}
\Big|^2 \\
 & & \hspace{-0.7cm}  c_{(s,1)}^{(n)}
\ \ \ = \ \ \frac{1}{N} \,\Big|
  + v_{(s,\gamma_5)}^{(n)}
  + v_{(c,\gamma_5)}^{(n)}
  + v_{(s,1)}^{(n)}
  - v_{(c,1)}^{(n)}
\Big|^2 \\
\label{EQN645} & & \hspace{-0.7cm} c_{(c,1)}^{(n)}
\ \ \ = \ \ \frac{1}{N} \,\Big|
  + v_{(s,\gamma_5)}^{(n)}
  + v_{(c,\gamma_5)}^{(n)}
  - v_{(s,1)}^{(n)}
  + v_{(c,1)}^{(n)}
\Big|^2 \ ,
\end{eqnarray}
where $N$ is a suitable normalization such that
\begin{eqnarray}
c_{(s,\gamma_5)}^{(n)}  + 
c_{(c,\gamma_5)}^{(n)}  + 
c_{(s,1)}^{(n)} + 
c_{(c,1)}^{(n)}  
 \ \ = \ \ 1 \ .
\end{eqnarray}
To give a specific example, if $c_{(c,\gamma_5)}^{(n)} \simeq 1$, 
while $c_{(s,\gamma_5)}^{(n)} \simeq c_{(s,1)}^{(n)} \simeq c_{(c,1)}^{(n)} \simeq 0$, 
the $n$-th state would be interpreted as the $D$ meson. In the continuum limit, where 
parity and heavy flavor symmetry are restored, each state will have one 
associated coefficient $c_{(h,\Gamma)}^{(n)}=1$, and all others vanishing.

\begin{figure}[htb]
\begin{center}
\input{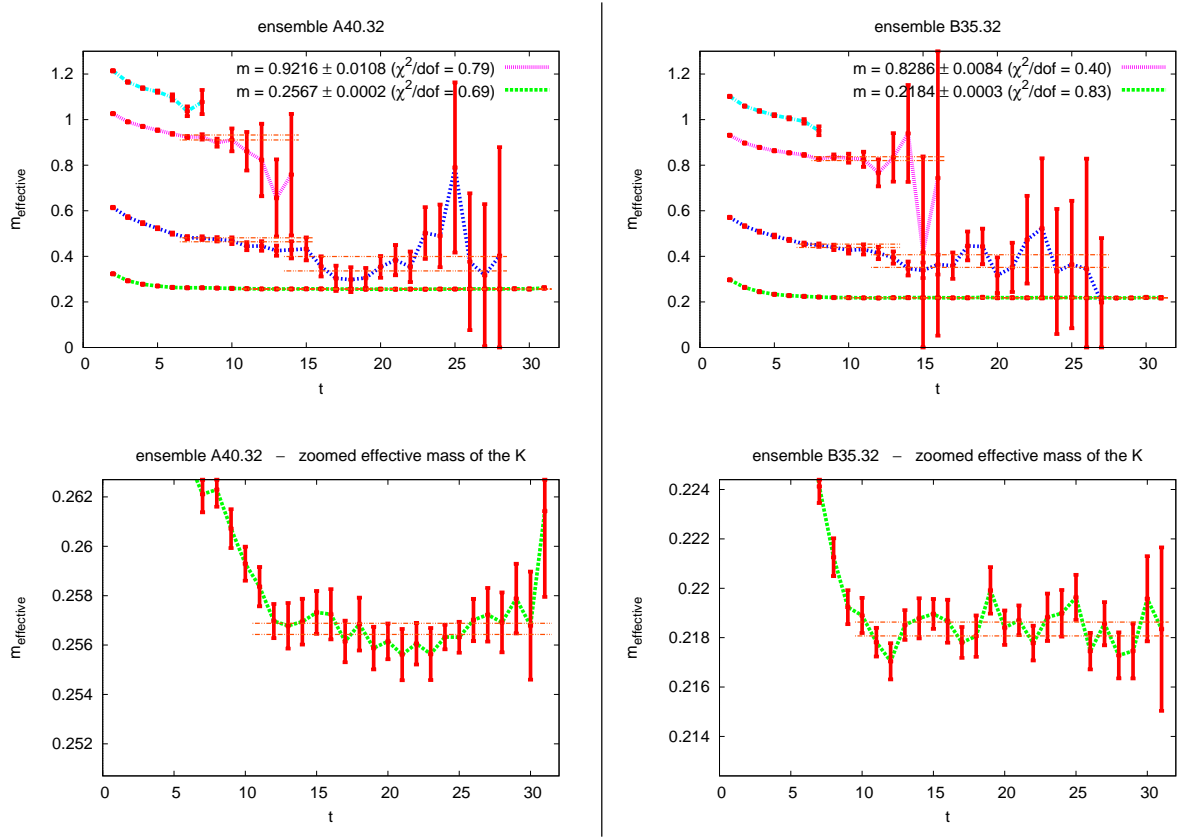}
\caption{\label{FIG003a} The four effective masses 
$m_\textrm{effective}^{(n)}$ as functions of $t$ ($t_0 = 1$) for the ensemble A40.32 (left) and B35.32 (right). The zoomed in effective masses for the $K$ meson are also shown in the bottom graphs.}
\end{center}
\end{figure}

\begin{figure}[htb]
\begin{center}
\input{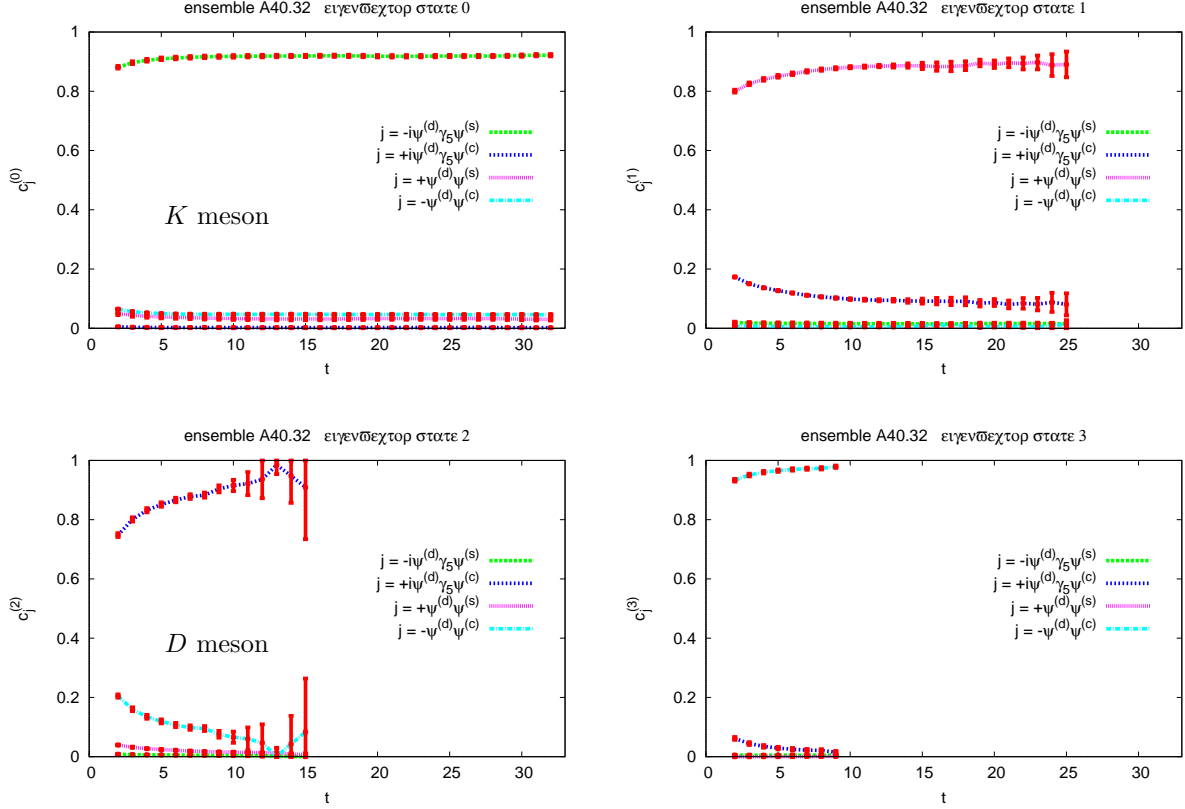}
\caption{\label{FIG003b} Approximate flavor and parity content of the four extracted 
states as a function of $t$ ($t_0 = 1$) for the ensemble A40.32. Top left: $n=0$, mainly $(s,-)$, 
i.e.\ the $K$ meson. Top right: $n=1$, mainly $(s,+)$. Bottom left: 
$n=2$, mainly $(c,-)$, i.e.\ the $D$ meson. Bottom right: $n=3$, mainly $(c,+)$.
The time ranges are the same as for the corresponding effective masses shown in 
Figure~\ref{FIG003a}.}
\end{center}
\end{figure}

Figure~\ref{FIG003a} shows the first four effective masses $m_\textrm{effective}^{(n)}$ 
($n = 0,\ldots,3$) as 
functions of $t$ for the ensembles A40.32 (left) and B35.32 (right), while 
 Figure~\ref{FIG003b} shows the approximate heavy flavor and parity contents of those states
for the ensemble A40.32, 
measured by the coefficients in eqs.~(\ref{EQN644}) to (\ref{EQN645}).
 As expected, each one of the effective masses is strongly dominated 
by and, therefore, should correspond to one of the sectors $(s,-)$, $(s,+)$, $(c,-)$ and 
$(c,+)$, which are approximately projected by the pseudo physical basis operators 
associated to the labels $(s,\gamma_5)$, $(s,1)$, $(c,\gamma_5)$ and $(c,1)$, respectively. 

To extract the numerical values for $m_K$ and $m_D$, we perform $\chi^2$ minimizing fits 
%%of constants 
to the corresponding effective mass plateaus. The fitting intervals 
$[t_\textrm{min}, t_\textrm{max}]$  are chosen as follows:
\begin{itemize}
\item $t_\textrm{max} = T/2 - 1 = 31$ for the $K$ meson.

\item For all the other states $t_\textrm{max}$ is the largest $t$ before which 
the corresponding effective mass is lost in statistical noise (cf.\ Table~\ref{TAB002}).

\item $t_\textrm{min}$ is the smallest $t$ fulfilling the following two requirements:
\begin{itemize}
\item $t_0 + 1 \leq t_\textrm{min} \leq t_\textrm{max}$.

\item All fitting intervals $[t_\textrm{min},t'_\textrm{max}]$, with 
$t_\textrm{min}+1 \leq t'_\textrm{max} \leq t_\textrm{max}$, 
yield a $\chi^2 / \textrm{dof} \leq (\chi^2 / \textrm{dof})_\textrm{max}$, and
 we require $(\chi^2 / \textrm{dof})_\textrm{max} = 2.0$.
\end{itemize}
By choosing $t_\textrm{min}$ in this way we prevent that effective masses at large $t$ 
with large statistical errors effectively increase the number of degrees of freedom, 
while not contributing to the $\chi^2$; in practice, 
the inclusion of these points would allow to fit ranges with too small values of 
$t_\textrm{min}$, outside the plateau region.
\end{itemize}
Within this method, a systematic error is associated to the determination of the $D$ meson 
mass, 
due to the fact that the effective mass plateau of the $(c,\,-)$ dominated state will finally
decay to lighter strange states at large times, as a consequence of
 the heavy flavor and parity breaking.
%Due to heavy flavor and parity breaking the effective mass plateaux of the $D$ meson 
%will finally break down to lighter strange states. Consequently, there is a systematic
% error involved in our determination of $m_D$. 

We account for this error by taking the difference with a fit
in the range $[t_\textrm{min} - 1, t_\textrm{max}]$, and we combine statistical and 
systematic uncertainties in quadrature, where the statistical error is obtained by 
a standard Jackknife analysis.
% We add this error in quadrature, i.e.\ the total error we quote is $\Delta_\textrm{total}
% = \sqrt{(\Delta_\textrm{statistical})^2 + (\Delta_\textrm{systematic})^2}$ (the statistical
% error is obtained by a standard Jackknife analysis).

The results for $m_K$, $m_D$ and the $(s,\, +)$ state, which for brevity
we denote from now on as $K_0^\ast$, are collected in Table~\ref{TAB002}.
\begin{table}[htb]
\begin{center}
\begin{tabular}{|c|r|c||c|r|c||c|r|c|}
\hline
 & & & & & & & & \vspace{-0.40cm} \\
$a m_K $ & $t$ range & $\chi^2 / \textrm{dof}$ & 
$a m_{K_0^\ast} $ & $t$ range & $\chi^2 / \textrm{dof}$ & $a m_D $ & $t$ range & 
$\chi^2 / \textrm{dof}$\vspace{-0.40cm} \\
 & & & & & & & & \\
\hline
\multicolumn{9}{|c|}{\vspace{-0.40cm}} \\
\multicolumn{9}{|c|}{Ensemble A40.32} \\
\multicolumn{9}{|c|}{\vspace{-0.40cm}} \\
\hline
 & & & & & & & & \vspace{-0.40cm} \\
$0.2567(2)$ & $11 - 31$ & $0.69$ & $0.368(32)$ & $14 - 28$ & $0.92$ & $0.922(11)$ & $7 - 14$ & $0.79$ \\
 & & & $0.473(15)$ & $\phantom{0}7 - 15$ & $1.65$ & & & \vspace{-0.40cm} \\
 & & & & & & & & \\
\hline
\multicolumn{9}{|c|}{\vspace{-0.40cm}} \\
\multicolumn{9}{|c|}{Ensemble B35.32} \\
\multicolumn{9}{|c|}{\vspace{-0.40cm}} \\
\hline
 & & & & & & & & \vspace{-0.40cm} \\
$0.2184(3)$ & $10 - 31$ & $0.83$ & $0.379(28)$ & $12 - 27$ & $0.54$ & $0.829(8)\phantom{0}$ & $8 - 16$ & $0.40$ \\
 & & & $0.446(7)\phantom{0}$ & $\phantom{0}7 - 13$ & $1.55$ & & & \vspace{-0.40cm} \\
 & & & & & & & & \\
\hline
\end{tabular}
\caption{\label{TAB002} The masses of the $K$, $K_0^\ast$ and $D$ mesons 
in lattice units obtained by solving a generalized eigenvalue problem (errors comprise statistical and systematic errors, which are added in quadrature). The range and the 
quality of the fit is also shown.}
\end{center}
\end{table}
As can also be inferred from Figure~\ref{FIG003a}, we obtain excellent results for 
$m_K$. For both ensembles the effective mass plateaus extend over 
more than twenty points, their statistical errors are essentially independent of $t$ and 
the relative errors on $m_K$ are $\approx 10^{-3}$.
For $m_D$ the situation is more problematic. As shown in Figure~\ref{FIG003a}, the 
corresponding effective masses are soon lost in statistical noise, before they reach 
unambiguously identifiable plateaus. As mentioned above, we add for this 
a systematic uncertainty.
The dominantly $(s,\,+)$ state does not exhibit a true plateau either. 
One rather observes two different plateaus, and we thus list two results for 
$m_{K_0^\ast}$ in Table~\ref{TAB002}, corresponding to two different fitting ranges. 
A possible explanation might be that at small temporal separations $t \ltapprox 10$ a 
positive parity strange meson is seen, while at larger $t$ the lighter $K + \pi$ state, with
the same strong quantum numbers, dominates. This is also supported by the fact that at 
larger values of the light quark mass a single plateau of rather good quality is recovered, see also the results in section~\ref{SEC422}.

%%%%%%%%%%%%%%%%%%%%%%%%%%%%%%%%%%%%%%%%%%%%%%%%%%%%%%%%%%%%%%%%%%%%%%%%%%%%%%%%%%%%%%%%%%%%%%%%%%%%%%

\subsection{\label{sec:method_2}Method 2: fitting the correlation matrix by exponentials}

A complementary approach to determine the heavy-light meson masses
is to fit the elements of the correlation 
matrix  of eq.~(\ref{EQN003}) by decomposing them in terms of the
 eigenstates of the Hamiltonian (i.e.\ the transfer matrix). 
%As detailed in section~\ref{sec:d_f_structure} we are concerned with a 
%$D \times D$ correlator matrix $C(t_2-t_1)$ (cf.\ (\ref{EQN003})). 
%It is defined by the matrix elements of $D$ operators 
%${\cal O}_\chi^{(a)},{\cal O}_\chi^{(b)},\ldots,{\cal O}_\chi^{(d)}$. 
We consider here the general case with different smearing levels,
where  $C(t_2-t_1)$, defined in eq.~(\ref{EQN003}), is a $D\times D$ matrix. 
%When including suitable factors $\pm i$ in the definition of the operators, 
%as it has been done in 
%(\ref{EQN001}), $C$ will be real and symmetric. 
When denoting the energy eigenstates by $|n\rangle$, $n=1,2,\ldots,M$, 
% then in a shorthand notation
% %
% \begin{eqnarray}
% \label{fit01} C(t_2-t_1) \ \ = \ \ \left(
% \begin{array}{cccc}
% C_{a,a}(t_2-t_1) & C_{a,b}(t_2-t_1) & \ldots & C_{a,d}(t_2-t_1) \\
% C_{b,a}(t_2-t_1) & C_{b,b}(t_2-t_1) & \ldots & C_{b,d}(t_2-t_1) \\
% \vdots          & \vdots          & \ldots & \vdots          \\
% C_{d,a}(t_2-t_1) & C_{d,b}(t_2-t_1) & \ldots & C_{d,d}(t_2-t_1)      
% \end{array}
% \right) ,
% \end{eqnarray}
% %
the matrix elements of  $C(t_2-t_1)$ can be written as
\begin{eqnarray}\label{eq:sd}
\label{fit02} C_{ij}(t_2-t_1) \ \ = \ \ \sum_{n=1}^{M}\ (i|n)_{t_2}\, (j|n)_{t_1} 
\end{eqnarray}
%
%%
%\begin{eqnarray}
%\label{fit02} C_{ij}(t_2-t_1) \ \ = \ \ (i|1)_{t_2}(j|1)_{t_1} + (i|2)_{t_2}(j|2)_{t_1} 
%+ \ldots + (i|M)_{t_2}(j|M)_{t_1}
%\end{eqnarray}
%%
with
\begin{eqnarray}
\label{fit03} (i|n)_t \ \ \equiv \ \ \langle \Omega | {\cal O}_\chi^{(i)}(t) | n \rangle \ \ 
= \ \ \langle n | \Big({\cal O}_\chi^{(i)}(t)\Big)^\dagger | \Omega \rangle \ ,
\end{eqnarray}
where $i=1, \ldots, D$ labels the operators 
inserted in the correlation matrix and $n=1,\ldots,M$ counts the eigenstates. 
Since we consider bosonic operators, we have a periodic time dependence on
the time extension of the lattice $T$ that can be written as follows
\begin{eqnarray}
\label{fit04} (i|n)_{t_2}\, (j|n)_{t_1} \ \ = \ \ (i|n) (j|n) \Big(\exp(-(t_2-t_1) E_n)+\exp(-(T-t_2+t_1) E_n)\Big)\, .
\end{eqnarray}
Here, $E_n$ is the energy of the eigenstate $|n\rangle$ 
and $(i|n) \equiv (i|n)_0$.
In general, the number of energy eigenstates is as large as the dimension of the Hilbert 
space of 
states. However, for large temporal separations $t_2-t_1,(T-t_2+t_1) \gg 1$ a few lowest
energy states will dominate to a good approximation. 
In this limit, and in analogy with the case of fitting a single correlation function
with the contributions from a few states, one can fit the matrix of correlation functions
with the contributions from the set of dominant 
lowest energy states. In fact, the relevant number of energy eigenstates $M$ is 
small. 
The number $N_P$ of parameters in the fit and the number $N_C$ of independent entries of 
$C(t_2-t_1)$ to be fitted are given by
\begin{eqnarray}
\label{fit05} N_P \ \ = \ \ M (D+1) \quad , \quad N_C \ \ = \ \ (t_\textrm{max}-t_\textrm{min}+1)\,\frac{D(D+1)}{2}\, ,
\end{eqnarray}
where also here $t_\textrm{min}$ and $t_\textrm{max}$ define the 
fitting time interval, with  $(t_2-t_1) \in [t_\textrm{min}, t_\textrm{max}]$.
The minimal set of operators for determining the heavy-light 
meson masses is given in this case by the $4 \times 4$ correlation matrix
%%%Istvan
in terms of the operators in
eq.~(\ref{EQN001}).
%
% \begin{equation}\label{fit06}
% Ps \equiv (\overline{\psi}^{(d)}\gamma_5 \psi^{(s)}) \ , \hspace{2em}
% Pc \equiv (\overline{\psi}^{(d)}\gamma_5 \psi^{(c)}) \ , \hspace{2em}
% Ss \equiv (\overline{\psi}^{(d)} \psi^{(s)}) \ , \hspace{2em}
% Sc \equiv (\overline{\psi}^{(d)} \psi^{(c)}) \ .
% \end{equation}
%
The minimal set of states we are interested in 
consists of the $K$ and $D$ mesons. At finite lattice spacing, due to the heavy flavor 
and parity breaking, 
the $D$ meson is not stable and does not correspond 
to an energy eigenstate of the lattice theory. However, 
using the same arguments of section~\ref{SEC003}, 
the $D$ should dominate the spectral decomposition of eq.~(\ref{eq:sd}) 
at intermediate temporal separations.
%stable
%However, below the $D$ meson there are other lighter states, 
%for instance $K + \pi$, $K + 2 \times \pi$, etc.\ states and kaonic states with positive parity. 
%As a consequence of the flavor and parity breaking of the lattice action there are non-vanishing
%matrix elements between the $D$ meson and these lower states and, therefore, the $D$ meson is not
%stable.
%This makes the determination of the $D$ meson mass difficult.
%(There is no difficulty in case of the kaon mass because the kaon is the lightest state and, 
%therefore, stable.) 
%%
%{\bf Elisabetta}
In case of fitting the correlation matrix by several exponentials, essential contributions 
of the lower sectors in the $D$ meson channel can be monitored by
considering the scalar products of the linear combination of the operators obtained from the fit
with the rows of the maximal twist matrix.
%{\bf Elisabetta}

Using the pseudo physical basis operators of eq.~(\ref{eq:op_phys_ren_mt}),
 one obtains for the coefficients $(i|n)$: 
\begin{eqnarray}
\label{eq:coeff_dec}
(i|n) \ \ \equiv \ \  \langle \Omega | {\cal O}_\chi^{(i)}(0) | n \rangle
\ \ = \ \  \sum_{j}\, \Big(\mathcal{M}_\textrm{mt}\Big)_{ji}\,
\langle \Omega | {\cal O}_\textrm{ppb}^{(j)}(0) | n \rangle\ .
\end{eqnarray}
Again, assuming that  $\omega_l =\omega_h \approx \pi / 2$ and $Z_P / Z_S=1$ are 
approximately verified, the operators ${\cal O}_\textrm{ppb}$ should reproduce the 
physical operators associated to the four channels $(s,-)$, $(s,+)$, $(c,-)$, $(c,+)$ 
to a good approximation. 
In particular, the operator with the same quantum numbers of the state
$|n\rangle$ should  dominate the sum in (\ref{eq:coeff_dec}). 
We therefore conclude
\begin{eqnarray}
(i|n) \ \ \simeq \ \  G_n \, \Big(\mathcal{M}_{\textrm{mt}}\Big)_{ni} \, 
\end{eqnarray}
to a good approximation, where the proportionality constant $G_n$ is
the matrix element of the physical operator:
\begin{eqnarray}
 G_n\ \ \equiv \ \ \langle \Omega | {\cal O}_\textrm{ph}^{(n)}(0) | n \rangle\ .
\end{eqnarray}
It turns out that it is enough to require that the 
relative signs of the vector 
components $(i|n)$ agree with the signs in the rows of
maximal twist rotation matrix $\mathcal{M}_{\textrm{mt}}$.
%{\bf Elisabetta} 
(A more stringent condition on the alignment with the rows of the maximal twist matrix could be 
imposed by requiring the scalar products of the linear combinations of the operators obtained 
from the fit with the rows of the maximal twist matrix to be close to 1,
but such a requirement does not essentially change the results for the $D$ meson mass.)
%{\bf Elisabetta}

Based on the experience with varying the number of states,  
we determine the $K$ meson mass with a single intermediate state, while 
good fits for the $D$ meson mass can be obtained  
for time separations around $t_2-t_1 \simeq 10-12$, by
using three intermediate states. Taking four states gives compatible 
results, but the signal is lost at smaller distances with consequently larger 
errors.
Larger correlation matrices have  also been investigated, for instance, 
$8 \times 8$ matrices spanned by four Gaussian smeared operators of type (\ref{EQN001}) 
and the corresponding four local operators. 
In this case stable fits with one, three or four states can also be obtained.

\begin{figure}[htb]
\begin{center}
\input{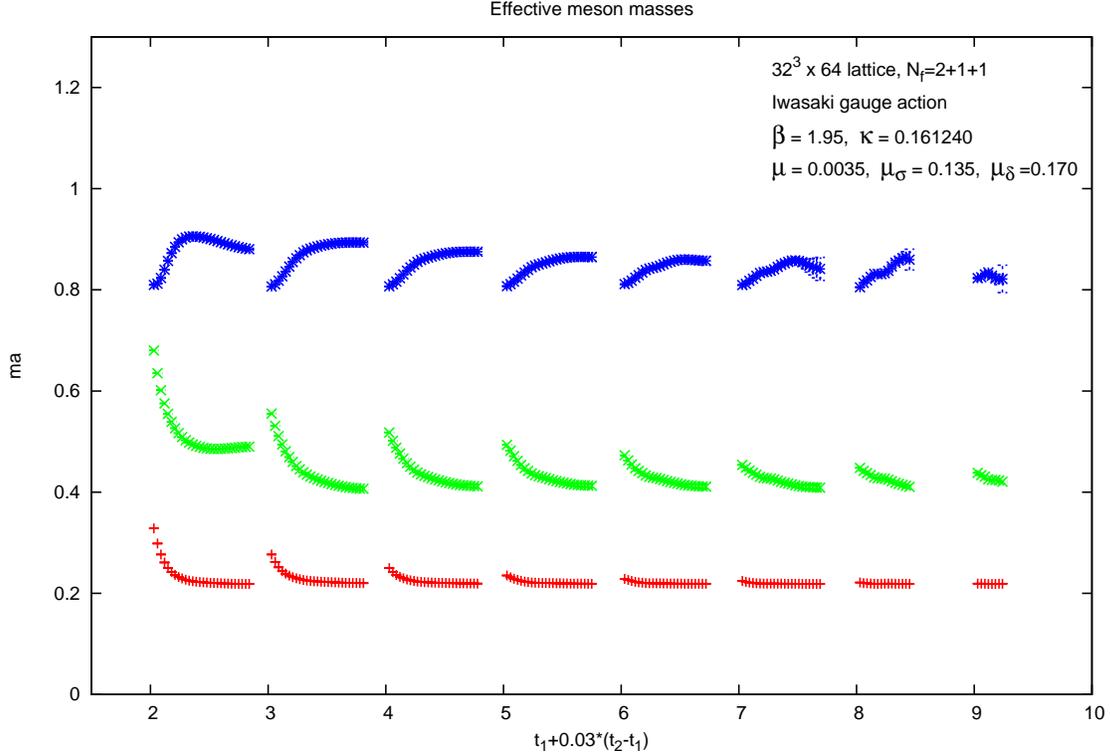}
\caption{\label{fitfig01}
Masses for the $K$, (bottom), $K_0^\ast$ (middle) and $D$ (top) channels 
obtained from a $3 \times 4$ fit, i.e.\ a $4 \times 4$ matrix with 3 states,
for the ensemble $B35.32$. 
The fit interval is $[t_1,t_2]$, with values shown by the abscissa. 
Errors on the masses are also plotted but in most 
cases are within the symbol size -- as show by the figure.}
%{\bf Elisabetta}
\end{center}
\end{figure}

We minimize the uncorrelated $\chi^2$
\begin{eqnarray}
\label{fit07} \chi^2 \ \ = \ \ \sum_{i=1}^{N_C} \bigg(\frac{f_i(p_1,p_2,\ldots,p_{N_P}) 
- \overline{X}_i}{\delta X_i}\bigg)^2 ,
\end{eqnarray}
where the index $i$ runs over the independent matrix elements to be fitted, 
$\overline{X}_i$ and 
$\delta X_i$ are the mean value and the error of the matrix element $i$ respectively, 
and $f_i(p_1,p_2,\ldots,p_{N_P})$ is 
the fitting function depending on $N_P$ parameters defined by 
eqs.~(\ref{fit02}) to (\ref{fit04}). 
We determined the errors of the matrix elements $\delta X_i$ and 
of the fit parameters $\delta p_i$ by the method in ref.~\cite{WOLFF}. 
Figure~\ref{fitfig01} illustrates how the extracted masses depend on the fit intervals.
We also studied the correlated $\chi^2$ following refs.~\cite{MICHAEL,Michael:1994sz}
\begin{eqnarray}
\label{fit08}\chi_c^2 \ \ = \ \ \sum_{i,j=1}^{N_C} 
\Big(f_i(p) - \overline{X}_i\Big) M_{ij} \Big(f_j(p) - \overline{X}_j\Big) ,
\end{eqnarray}
where $M_{ij} = N \mathcal{C}_{ij}^{-1}$, with $N$ input data and the estimated 
covariance matrix
\begin{eqnarray}
\label{fit09} \mathcal{C}_{ij} \ \ = \ \ \frac{1}{N-1}\sum_{n=1}^{N} 
\Big(X_{i,n} - \overline{X}_i\Big) \Big(X_{j,n} - \overline{X}_j\Big) .
\end{eqnarray}
It turned out, however, that on our data samples the covariance matrix has a large number of 
almost degenerate tiny eigenvalues of the order of magnitude $10^{-16}$,
%{\bf Elisabetta}
which cannot be properly 
determined within the present statistical accuracy. The small eigenvalues can be smoothed  
\cite{MICHAEL,Michael:1994sz}, at the price of introducing an uncertainty in the value 
of $\chi_c^2$. 
For this reason, we decided to minimize the uncorrelated $\chi^2$, 
and to use the correlated one $\chi_c^2$ to estimate systematic errors, see below.

Relative errors of the elements of our correlation matrices are typically of 
 ${\cal O}(10^{-2})$. This results in rather small errors for the fit parameters on
 a given time interval: masses have relative errors of ${\cal O}(10^{-2})$ 
to ${\cal O}(10^{-3})$, while the components of the energy eigenvectors have errors 
${\cal O}(10^{-2})$. 
A good fit has to satisfy for our case the following requirements:
\begin{enumerate}
\item 
The quantum number pattern of the fitting operators has to be as expected, 
i.e. the relative signs of the components of the fitted vectors are the same as those of the 
rows of the maximal twist matrix.

\item 
We exclude the results from fit intervals, where the relative errors of the masses are 
substantially higher than the typical errors. With our statistics, this means 1\% for the 
$K$ meson mass and 5\% for the other masses. Only a few fit intervals turn out to be
 affected by this choice.

\item
The fit ranges $[t_{min},t_{max}]$ are restricted by applying cuts in $t_{min}$ and 
$(t_{max}-t_{min})$ 
%{\bf Elisabetta}
such that a reasonable ``plateau'' of the fit values emerges, 
always keeping a sufficiently large number of fit ranges in the sample,
typically about 30 to 80.
%{\bf Elisabetta}
\end{enumerate}
%%
%The deviations between the results of ``good fits'' are comparable and, therefore, this 
%spread of the results has to be taken into account in the final error. In other words, 
%the systematic error due to the choice of the fitting interval has to be estimated.

After selecting a set of good fits by these criteria a histogram distribution of the fit 
values has been defined by attributing a weight $\exp(-\chi_c^2/\textrm{dof})$ to the 
entries in case of the kaon, and a weight $1/(\chi_c^2/\textrm{dof})$ in the other channels.
The exponential suppression is in general preferable, since 
 it gives robust results but can only 
be applied for very good fits and plateaus, which is the case for the kaon.
In order to combine statistical and systematic errors, the entries in the distribution 
were not attributed to a single point but uniformly to the points on the interval
$[p_i-\delta p_i,p_i+\delta p_i]$. 
For each final quantity, 
the quoted value is then the position of the median of the
resulting distribution. The error is given by a symmetric interval around the 
median such that 68\% of the distribution is contained in it.
%{\bf Elisabetta}
%{\bf Elisabetta}

We report on single-state, three-state, and 
four-state fits with a $4 \times 4$ correlation matrix of Gaussian smeared operators. 
For completeness, we also show the results of three-state fits with an 
$8 \times 8$ matrix of 
Gaussian smeared and local operators. 
All results are summarized in Table~\ref{fittable01}.

%%%%%%%%%%%%%%%%%%%%%%%%%%%%%%%%%%%%%%%%%%%%%%%%%%%%%%%%%%%%%%%%%%%%%%%%%%%%%%%%
\begin{table}[htb]
\begin{center}

\begin{tabular}{|c*{6}{|c}|}
\hline
& & & & & \vspace{-0.40cm}
\\ 
Ensemble & $M\times D$ & $am_K$ & $am_{K_0^\ast} $ & $am_D $ & $a m_{D_0^\ast} $
\\& & & & & \vspace{-0.40cm}
\\ \hline
A40.32  & 1x4  & 0.25542(67)  &            &            &
\\ \hline
     &    3x4  & 0.25853(88)  & 0.448(13)  & 0.903(20)  &
\\ \hline
    &  4x4  & 0.26272(62)  & 0.4905(60) & 0.939(46)  & 1.09(15)
\\ \hline
    & 3x8  & 0.2627(23)   & 0.478(18)  & 0.885(22)  &
\\ \hline\hline
B35.32 & 1x4  & 0.21766(64)  &            &            &
\\ \hline
   & 3x4  & 0.21864(51)  & 0.422(11)  & 0.835(20)  &
\\ \hline
   & 4x4  & 0.2226(69)   & 0.449(24)  & 0.896(70)  & 1.19(10)
\\ \hline
   & 3x8  & 0.2203(13)   & 0.4369(94) & 0.814(18)  &
\\ \hline
\end{tabular}
%%%
\caption{\label{fittable01} Masses of the $K$, $K_0^\ast$, $D$ and $D_0^\ast$ in 
lattice units, resulting from the fits to the correlation matrices with 
several eigenstates for 
the ensembles A40.32 and B35.32. The label $M \times D$ means a fit with $M$ eigenstates 
to a $D \times D$ matrix.}
\end{center}
\end{table}
%%%%%%%%%%%%%%%%%%%%%%%%%%%%%%%%%%%%%%%%%%%%%%%%%%%%%%%%%%%%%%%%%%%%%%%%%%%%%%%%

As shown in table~\ref{fittable01} the four-state fit to a $4 \times 4$ matrix 
gives one state in each of 
the channels $J^\mathcal{P} = 0^-$ and $J^\mathcal{P} = 0^+$, with both strange and 
charmed quarks. On the other hand, errors are typically larger and/or the light states have 
higher masses than in the $1 \times 4$ and $3 \times 4$ fits. 
Therefore, as final results we 
quote the $K$ meson mass from the $1 \times 4$ fit and the $D$ meson mass from the 
$3 \times 4$ fit.

One can verify a posteriori how 
well the quantum number content of each fitted vector corresponds to the expected one. 
This is simply given by the scalar product of the unit vector in the direction of the 
fitted vector with the row of the matrix in eq.~(\ref{eq:op_phys_ren_mt}) that 
gives the expected vector in the continuum limit at maximal twist. 
For this, we remind that the $K$ meson, strange $0^+$ state, $D$ meson and 
charmed $0^+$ state correspond 
to the rows 1, 3, 2 and 4, respectively. 
Table~\ref{fittable02} shows that the fitted vectors are actually well 
saturated by the expected quantum numbers, with scalar products 
close to 1 in all cases.
%%%%%%%%%%%%%%%%%%%%%%%%%%%%%%%%%%%%%%%%%%%%%%%%%%%%%%%%%%%%%%%%%%%%%%%%%%%%%%%%
\begin{table}[ht]
\begin{center}

\begin{tabular}{|c*{6}{|c}|}
\hline
& & & & & \vspace{-0.40cm}
\\Ensemble & $M\times D$ & $z_K$ & $z_{K_0^\ast}$ & $z_D$ & $z_{D_0^\ast}$
\\ & & & & &\vspace{-0.40cm}
\\ \hline
A40.32 & 1x4     & 0.98659(6)   &             &           &
\\ \hline
&3x4     & 0.9871(2)    & 0.9896(17)  & 0.9392(78)&
\\ \hline
&4x4     & 0.9870(2)    & 0.9845(23)  & 0.9929(1) & 0.9830(133)
\\ \hline\hline
B35.32 & 1x4     & 0.98518(8)   &             &           &
\\ \hline
& 3x4     & 0.9847(1)    & 0.9772(33)  & 0.9518(94)&
\\ \hline
& 4x4     & 0.9848(1)    & 0.9770(21)  & 0.9777(86) & 0.9732(110)
\\ \hline
\end{tabular}

\caption{\label{fittable02} Saturation of the fitted states
with the expected quantum numbers, for the four-states fits of table \ref{fittable01},   
measured by the scalar product $z$ (see text). A value $z=1$ indicates complete 
saturation. Mean values and errors for $z$ are determined analogously to masses.}
\end{center}
\end{table}

%%%%%%%%%%%%%%%%%%%%%%%%%%%%%%%%%%%%%%%%%%%%%%%%%%%%%%%%%%%%%%%%%%%%%%%%%%%%%%%%%%%%%%%%%%%%%%%%%

% **********

\subsection{\label{SEC422}Method~3: parity and flavor symmetry restoration}

This third method is a generalization of the ``parity restoration method'' originally
 introduced for the twisted mass formulation with two 
degenerate quarks~\cite{Farchioni:2004ma,Farchioni:2004fs,Farchioni:2005bh}. 
In the $N_{\rm f}=2$ setup the twist angle can be determined by requiring that the operators
 reproducing the correct definition of the chiral currents in the continuum limit (physical 
chiral currents) possess the appropriate transformation properties under parity. 
This condition allows to fix the twist angle for the degenerate light quark doublet and 
the correctly normalized physical currents. We generalize the method to the case of 
 bilinear densities with mixed heavy-light flavor composition, used here 
for the determination of the $K$ and $D$ meson masses. A first account of this 
method can be found in~\cite{Chiarappa:2006ae}. As an outcome, approximations of the
physical operators in eq.~(\ref{EQN001_ph}) can be constructed,
from which the masses in the four heavy-flavor and parity channels can be extracted 
by conventional techniques.

Consider the four-by-four correlation matrix of the renormalized lattice operators in eq.~ 
(\ref{eq:vec_chi_ren}):
\begin{eqnarray}
\label{eq:corr_mat_r} C^R(t_2-t_1) \ \ = \ \ \Big\langle \mathcal{O}^R(t_2) \otimes (\mathcal{O}^R(t_1))^\dagger \Big\rangle \, .
\end{eqnarray}
After rewriting the renormalized lattice operators in terms of the bare ones one obtains 
\begin{eqnarray}
\label{eq:corr_mat_r_2} C^R(t_2-t_1) \ \ = \ \  \textrm{diag}\Big(Z_P,Z_P,Z_S,Z_S\Big)\, C(t_2-t_1)\, \textrm{diag}\Big(Z_P,Z_P,Z_S,Z_S\Big) \ ,
\end{eqnarray}
where $C(t_2-t_1)$ is the correlation matrix defined in eq.~(\ref{EQN003}), 
the starting point of the previous two methods. 
The transformation properties of the correlation matrix (\ref{eq:corr_mat_r})
can be read from eq.~(\ref{eq:op_phys_ren}), implying that the correlation matrix of the 
physical operators (\ref{EQN001_ph}) is given by
\begin{eqnarray} \nonumber
C_\textrm{ph}^R &=&  \mathcal{M}(\omega_l,\omega_h)\, C^R\, \mathcal{M}^T(\omega_l,\omega_h) \ = \ 
\\\label{eq:corr_phys}
&& 
{\mathcal M}(\omega_l,\omega_h)\, 
\textrm{diag}\Big(Z_P,Z_P,Z_S,Z_S\Big)\, C\, \textrm{diag}\Big(Z_P,Z_P,Z_S,Z_S\Big)
{\mathcal M}^T(\omega_l,\omega_h) \ ,
\end{eqnarray}
where, we recall, the general orthogonal twist rotation matrix ${\mathcal M}(\omega_l,\omega_h)$ 
is given by (\ref{eq:mix}).
Since we are working at maximal twist, we are supposed to insert $\omega_l=\omega_h=\pi/2$
in the rotation matrix of eq.~ (\ref{eq:corr_phys}). 
However, differently from the previous two methods and accounting for the presence of 
$\Oa$ effects, we 
treat the two twist angles, along with the renormalization 
factors $Z_P$ and $Z_S$, as free parameters. We will return to this point in the following. 
These free parameters can be determined by imposing that the physical 
operators indeed possess the appropriate parity and flavor quantum numbers
of their associated channel. This in particular implies that the physical correlation matrix 
of eq.~(\ref{eq:corr_phys}) should be
diagonal
\begin{eqnarray}
\Big(C^R_\textrm{ph}\Big)_{j k} \ \ = \ \ 0\ , \quad  j \neq k \ .
\label{EQN693}
\end{eqnarray}
Since $C(t_2-t_1)$ is a symmetric matrix (see section~\ref{sec:d_f_structure}), 
the matrix in eq.~(\ref{eq:corr_phys}) is by construction symmetric and
eq.~(\ref{EQN693}) actually amounts to only six independent conditions. The latter can be rearranged as follows
\begin{eqnarray}
\label{eq:zeta} & & \hspace{-0.7cm} \frac{Z^2_P}{Z^2_S} \ \ = \ \ -\frac{C_{3 4}}{C_{1 2}} \\
\label{eq:oml} & & \hspace{-0.7cm} \mathrm{ctg}(\omega_l) \ \ = \ \ +\frac{(+C_{1 1}-C_{2 2}) (Z_P/Z_S) + (-C_{3 3}+C_{4 4}) (Z_S/Z_P)}{2  (C_{1 3}-C_{2 4})} \\
\label{eq:omh} & & \hspace{-0.7cm} \mathrm{ctg}(\omega_h) \ \ = \ \ +\frac{(+C_{1 1}-C_{2 2}) (Z_P/Z_S) + (+C_{3 3}-C_{4 4}) (Z_S/Z_P)}{2  (C_{1 4}-C_{2 3})} \\
\nonumber & & \hspace{-0.7cm} \mathrm{tan}(\omega_l+\omega_h) \ \ = \\
\label{eq:oml_2} & & = \ \ -\frac{C_{14}+C_{23}+C_{13}+C_{24}}{(+C_{1 1}+C_{2 2}) (Z_P/Z_S) + (-C_{3 3}-C_{4 4}) (Z_S/Z_P))/2 + C_{12} (Z_P/Z_S) - C_{34} (Z_S/Z_P)} \\
\nonumber & & \hspace{-0.7cm} \mathrm{tan}(\omega_l-\omega_h) \ \ = \\
\label{eq:omh_2} & & = \ \ +\frac{C_{14}+C_{23}-C_{13}-C_{24}}{(+C_{1 1}+C_{2 2}) (Z_P/Z_S) + (-C_{3 3}-C_{4 4}) (Z_S/Z_P))/2 - C_{12} (Z_P/Z_S) + C_{34} (Z_S/Z_P)} \\
\label{eq:oml_omh} & & \hspace{-0.7cm} \frac{\mathrm{tan}(\omega_l)}{\mathrm{tan}(\omega_h)} \ \ = \ \ -\frac{C_{1 3}+C_{2 4}}{C_{1 4}+C_{2 3}} .
\end{eqnarray}
%
% \begin{eqnarray}
% \label{eq:zeta} & & \hspace{-0.7cm} \frac{Z^2_P}{Z^2_S} \ \ = \ \ \frac{\langle S_{K^+} S_{\bar D^0} \rangle + \langle S_{D^0} S_{K^-} \rangle}{\langle P_{K^+} P_{\bar D^0} \rangle + \langle P_{D^0} P_{K^-} \rangle} \\
% 
% \label{eq:oml} & & \hspace{-0.7cm} \mathrm{ctg}(\omega_l) \ \ = \ \ \frac{C_{11}+C_{44}-C_{22}-C_{33}}{2(C_{24}-C_{13})} \\
%
% \label{eq:omh} & & \hspace{-0.7cm} \mathrm{ctg}(\omega_h) \ \ = \ \ \frac{C_{11}+C_{33}-C_{22}-C_{44}}{2(C_{23}-C_{14})} \\
%
% \label{eq:oml_2} & & \hspace{-0.7cm} \mathrm{tan}(\omega_l+\omega_h) \ \ = \ \ \frac{C_{14}+C_{23}+C_{13}+C_{24}}{(C_{33}+C_{44}-C_{11}-C_{22})/2-(C_{12}-C_{34})} \\
%
% \label{eq:omh_2} & & \hspace{-0.7cm} \mathrm{tan}(\omega_h-\omega_l) \ \ = \ \ \frac{C_{14}+C_{23}-C_{13}-C_{24}}{(C_{33}+C_{44}-C_{11}-C_{22})/2+C_{12}-C_{34}} \\
%
% \label{eq:oml_omh} & & \hspace{-0.7cm} \frac{\mathrm{tan}(\omega_l)}{\mathrm{tan}(\omega_h)} \ \ = \ \ \frac{C_{13}+C_{24}}{C_{14}+C_{23}} .
% \end{eqnarray}
%
Observe that the right hand sides of (\ref{eq:oml}) to (\ref{eq:oml_omh}) are fully 
determined by the ratio $Z_P/Z_S$, i.e.\ they do not depend individually on either $Z_P$ or $Z_S$. 

\begin{figure}[htb]
\begin{center}
\input{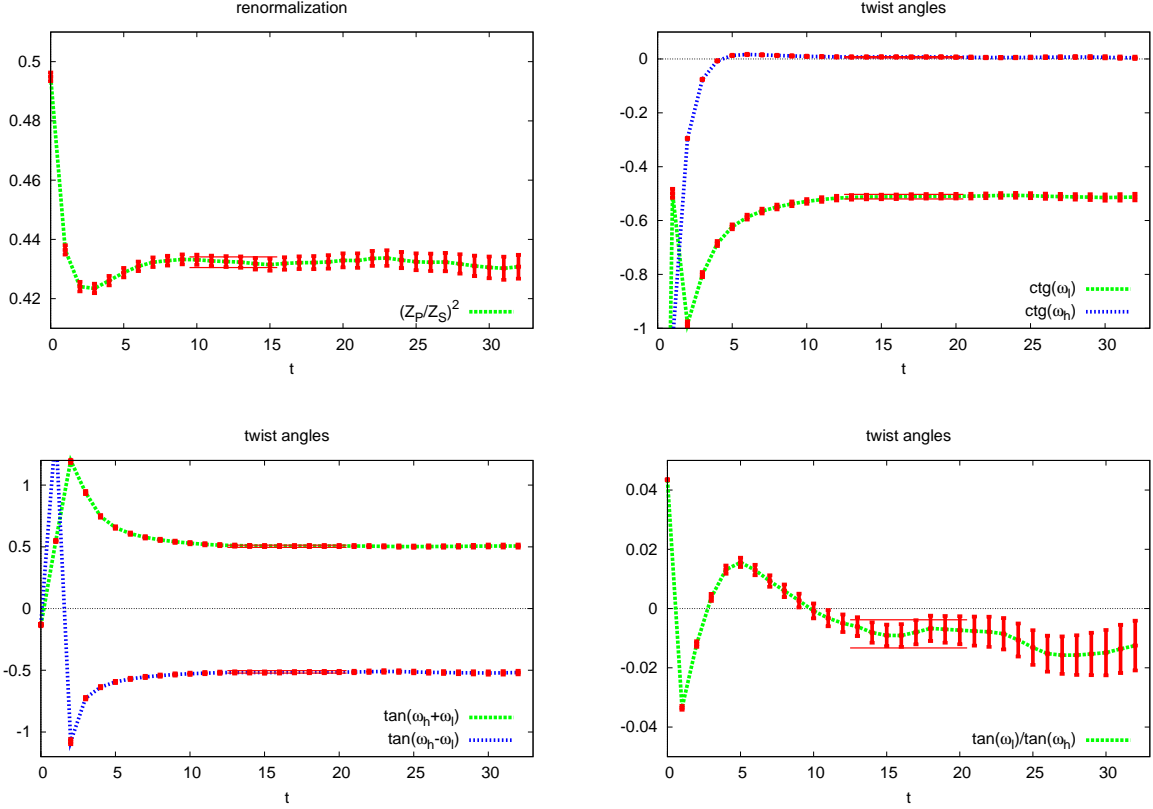}
\caption{\label{fig:parameters_A}ratios of correlators corresponding to the right hand sides of the conditions (\ref{eq:zeta}) to (\ref{eq:oml_omh}) as functions of the temporal separation $t$ for ensemble A40.32 with point-like operators (i.e.\ no Gaussian smearing); the lines indicate the fits in the asymptotic regime.}
\end{center}
\end{figure}

In Figure~\ref{fig:parameters_A} we report on the ratios of correlators on the right hand 
sides of the conditions (\ref{eq:zeta}) to (\ref{eq:oml_omh}) as functions of the time 
separation $t\equiv t_2-t_1$, for the ensemble A40.32 and the original operators without
Gaussian smearing. 
The ratios appear to approach a plateau after a transient: from these plateaus we determine the unknown parameters  $Z_P/Z_S$, $\omega_l$ and $\omega_h$. 

Notice that the time dependence of the ratios 
is an $\mathcal{O}(a)$ discretization effect, and
therefore not predicted by eqs.~(\ref{eq:zeta})-(\ref{eq:oml_omh}), which were derived in 
the continuum limit.
For large times the lightest eigenstate of the lattice transfer matrix, corresponding to 
the kaon in the continuum, is supposed to saturate the spectral decomposition of the 
correlation matrix $C(t_2-t_1)$ (see eqs. (\ref{eq:sd}) and (\ref{fit03})). 
Assuming a single intermediate state, 
the six conditions (\ref{eq:zeta})-(\ref{eq:oml_omh}) are not independent any more and
in particular the first three of relations (\ref{eq:zeta})-(\ref{eq:omh}) are equivalent to
(\ref{eq:oml_2})-(\ref{eq:oml_omh}).
Parity and flavor restoration amounts in this case to requiring that the three physical 
operators associated to the heavier channels have no projection on the lightest state, namely
\begin{eqnarray}
\label{eq:as_cond_me}
\sum_{\bf x}\, \langle  \Omega | {\mathcal{O}_\textrm{ph}^{(s,1)}} ({\bf x},t)| K \rangle \ \ = \ \ 
\sum_{\bf x}\, \langle  \Omega | {\mathcal{O}_\textrm{ph}^{(c,\gamma_5)}} ({\bf x},t)| K \rangle \ \ = \ \ 
\sum_{\bf x}\, \langle  \Omega | {\mathcal{O}_\textrm{ph}^{(c,1)}} ({\bf x},t)| K \rangle \ \ = \ \ 0\ .
\end{eqnarray}
We also observe that this procedure, which relies on asymptotic times, is supposed to be 
optimal from the point of view of the cutoff effects:
at large times, contribution from high-mass intermediate states, which are expected to introduce 
large discretization effects in the correlator, is suppressed.
A similar argument was used when tuning the 
theory to maximal 
twist in the light sector, see~\cite{Boucaud:2008xu}.

We determine $Z_P/Z_S$, $\omega_l$ and $\omega_h$ by using the relations 
(\ref{eq:zeta}-\ref{eq:omh}), 
while the remaining relations serve for cross checking of the results.
The latter are reported in Table~\ref{tab:par}. We observe an excellent agreement 
between the different determinations of the twist angles from 
(\ref{eq:oml}-\ref{eq:omh}) and (\ref{eq:oml_2}-\ref{eq:omh_2}), respectively,
confirming that a single intermediate state contributes. 
The quality of the agreement deteriorates, of course, when the parameters are estimated at smaller temporal separations outside the asymptotic region.
Notice that 
the ratio $\tan(\omega_l)/\tan(\omega_h)$ is in all cases compatible with zero, since 
$\omega_h \approx \pi/2$. 
%%%
\begin{table}[htb]
\begin{center}
\renewcommand{\arraystretch}{1.0}
\begin{tabular}{|c|c|c|c|l|c|}
\hline
 \multicolumn{1}{|c|}{eqs.}&
 \multicolumn{1}{|c|}{$Z_P/Z_S$} &
 \multicolumn{1}{|c|}{eqs.}&
 \multicolumn{1}{|c|}{$\omega_l/\pi$ } &
 \multicolumn{1}{|c|}{$\omega_h/\pi$ } &
 \multicolumn{1}{|c|}{$\tan(\omega_l)/\tan(\omega_h)$} 
 \\
\hline
% \multicolumn{6}{c}{$\beta=1.9$, $\mu=0.004$, $L_x=24$} \\
% \hline
%  (\ref {eq:zeta}) & 0.6588(15) & (\ref{eq:oml}-\ref{eq:omh})    & 0.6571(25) & 0.4990(8)  & -0.006(5)
%  \\ \hline
%                   &            & (\ref{eq:oml_2}-\ref{eq:omh_2})& 0.6571(25) & 0.4990(10) & -0.006(6)
%  \\ \hline 
%                   &            & (\ref{eq:oml_omh})             &            &            & -0.007(5)
%  \\ \hline\hline
% \multicolumn{6}{c}{$\beta=1.9$, $a\mu=0.004$, $L_x=32$ (set $A$)} \\
 \multicolumn{6}{c}{ensemble A40.32} \\
 \hline
 (\ref {eq:zeta}) & 0.6575(14)       & (\ref{eq:oml}-\ref{eq:omh})     & 0.6504(21) & 0.4980(8)  & -0.012(5)
 \\ \hline
 (\ref {eq:zeta}) & {\rm same value} & (\ref{eq:oml_2}-\ref{eq:omh_2}) & 0.6498(22) & 0.4990(10) & -0.006(5)
 \\ \hline 
 (\ref {eq:zeta}) & {\rm same value} & (\ref{eq:oml_omh})              & \multicolumn{1}{c|}{$-$} & \multicolumn{1}{c|}{$-$} &  -0.009(5)
  \\ \hline
%\multicolumn{6}{c}{$\beta=1.95$, $a\mu=0.0035$, $L_x=32$ (set $B$)} \\
 \multicolumn{6}{c}{ensemble B35.32} \\
\hline
 (\ref {eq:zeta}) & 0.6793(22)        & (\ref{eq:oml}-\ref{eq:omh})    &  0.6453(34) &  0.5005(8) &   0.003(5)      
 \\ \hline
 (\ref {eq:zeta}) & {\rm same value}  & (\ref{eq:oml_2}-\ref{eq:omh_2})&  0.6467(29) &  0.5007(9) &   0.005(6)      
 \\ \hline 
  (\ref {eq:zeta})& {\rm same value}  & (\ref{eq:oml_omh})             & \multicolumn{1}{c|}{$-$}            &  \multicolumn{1}{c|}{$-$}          &   0.005(5)      
 \\ \hline
\end{tabular}

\caption{\label{tab:par}summary of different determinations of the ratio of renormalization factors and of the twist angles with point-like operators (no Gaussian smearing); the first and third column indicate the equations used for the determination of the quantities in the corresponding line.}

\end{center}
\end{table}
Note instead that the value of the light twist angle in Table~\ref{tab:par} 
significantly deviates from the expected value $\pi/2$. 
In order to understand this discrepancy it is useful to 
recall that the theory is tuned to maximal twist by requiring the vanishing of the 
untwisted PCAC quark mass 
$m_{\chi^{(l)}}^{PCAC}$ {\em in the light quark sector}, see eq.~(\ref{eq:PCAC})). 
This can be shown to be equivalent~\cite{Farchioni:2005bh} to 
requiring parity restoration in the light quark sector. 
One constructs in this case the physical vector current as follows~\cite{Farchioni:2005bh}
\begin{eqnarray}
V^{(l) +}_{ph}(x) \ \propto\  \cos(\omega_l)\, Z_V\,V^{(l) +}(x)\,-i\,\sin(\omega_l)\, Z_A\,A^{(l) +}(x) \ ,
\end{eqnarray}
where the bilinear of the lattice fields $A^{(l) +}(x)$ is defined in 
eq.~(\ref{eq:PCAC_}) and, 
analogously,
\begin{eqnarray}
 V^{(l) +}_\mu \ \ = \ \ \bar\chi^{(u)} \gamma_\mu \chi^{(d)}\ ,
\end{eqnarray}
and $Z_A$, $Z_V$ are the respective renormalization constants in the massless scheme.  
The twist angle $\omega_l$ is fixed in this case by  the condition
\begin{eqnarray}\label{eq:cond_light}
\sum_{\bf x} \langle \Omega | V^{(l) +}_0 ({\bf x},t)| \pi \rangle \ =\ 0\ ,
\end{eqnarray}
from which one obtains
\begin{eqnarray}\label{eq:om_l_pi}
{\rm ctg}(\omega_l) \ \ = \ \ \frac{Z_A\,m_{\chi^{(l)}}^\textrm{PCAC}}{\mu} \ .
\end{eqnarray}
From this we can conclude that our maximal twist condition 
$m_{\chi^{(l)}}^\textrm{PCAC} = 0$ amounts to $\omega_l = \pi / 2$, 
if the condition (\ref{eq:cond_light}) is assumed.
This must be confronted with the conditions (\ref{eq:as_cond_me}) presently 
used to fix the twist angles $\omega_l$\footnote{In the asymptotic regime, where only 
the kaon state is considered as intermediate state, the light twist angle $\omega_l$ is 
fixed by the vanishing of the first two matrix elements in (\ref{eq:as_cond_me}); this is 
so because, in this regime, the two conditions can be proven to imply in particular 
relations (\ref{eq:zeta}) and (\ref{eq:oml}) (analogously, $\omega_h$ is fixed in 
particular by the vanishing of the second and third  matrix element).}. 
We conclude that the deviation of $\omega_l$ from $\pi/2$ should be attributed to 
different $\mathcal{O}(a)$ effects in the pion and kaon sectors.

We stress that the prescription of eq.~(\ref{eq:PCAC}), which is based on the charged 
pion state, is to be preferred for tuning the theory to maximal twist, since it ensures 
the smallest $\mathcal{O}(a^2)$ discretization errors in physical 
quantities~\cite{Frezzotti:2005gi}. Nevertheless, for the determination of the masses in 
the heavy-light meson sector, we use the values of the twist angles 
obtained from (\ref{eq:zeta}) to (\ref{eq:oml_omh}), since they deliver 
optimal projecting operators as defined in eq.~ (\ref{eq:op_phys_ren}),
 with the smallest heavy flavor and parity violations.
% Theoretically, neglecting in this case $\mathcal{O}(a)$ corrections, one expects
% %
% \begin{eqnarray}
% \label{eq:theo_oml} & & \hspace{-0.7cm} {\rm ctg}(\omega_l) \ \ = \ \ \frac{Z_A m_\chi^\textrm{PCAC}}{\mu} \\
% %
% \label{eq:theo_omh} & & \hspace{-0.7cm} {\rm ctg}(\omega_h) \ \ = \ \ \frac{Z_A m_\chi^\textrm{PCAC}}{\mu_\sigma} .
% \end{eqnarray}
% %
The relation in (\ref{eq:om_l_pi}) can also be enforced for the present determination of the light twist angle 
with heavy-light quark bilinears, and the cutoff effects can be absorbed in a 
lattice redefinition of the PCAC quark mass, 
$\tilde m_{\chi^{(l)}}^\textrm{PCAC} =  m_{\chi^{(l)}}^\textrm{PCAC} + \mathcal{O}(a)$. 
For the ensemble A40.32, we get for instance $Z_A\tilde m_{\chi^{(l)}}^\textrm{PCAC}/\mu 
\approx -0.5$, a pretty large value\footnote{For comparison, in the tuning procedure 
we require $Z_A |m_{\chi^{(l)}}^\textrm{PCAC}| / \mu \leq 0.1$.}. 
The analogous of relation (\ref{eq:om_l_pi}) for the heavy twist angle reads
\begin{eqnarray}
\label{eq:theo_omh} 
{\rm ctg}(\omega_h) \ \ = \ \ \frac{Z_A \tilde m_{\chi^{(l)}}^\textrm{PCAC}}{\mu_\sigma} \, .
\end{eqnarray}
The heavy twisted mass $\mu_\sigma$ replaces the light twisted mass $\mu$,  explaining 
why $\omega_h$ is very close to $\pi/2$: since $\mu_\sigma\gg\mu$, the non-zero value of 
$\tilde m_{\chi^{(l)}}^\textrm{PCAC}$ only results in a small deviation of $\omega_h$ from maximal twist. When inserting the above estimate in (\ref{eq:theo_omh}) we indeed obtain $\omega_h=0.4956$.

The ratio of normalization factors $Z_P/Z_S$ and the twist angles $\omega_l$ and $\omega_h$ 
allow to determine the physical operators up to an overall renormalization 
(bare physical operators). 
We choose this renormalization to be $Z_P$, so that (cf. eqs.~(\ref{eq:op_phys_ren}) 
and (\ref{eq:vec_chi_ren})) 
\begin{eqnarray}
\label{eq:bare_phys_corr} 
 \mathcal{O}_\textrm{ph}^{bare} \ \equiv\ Z_P^{-1}\, \mathcal{O}_\textrm{ph}^{R} \ = \ {\mathcal M}(\omega_l,\omega_h)\, 
\textrm{diag}\Big(Z_P/Z_S,Z_P/Z_S,1,1\Big)\,  \mathcal{O}_\chi\ .
\end{eqnarray}
Observe that in the case of the negative parity densities, eq.~(\ref{eq:bare_phys_corr}) 
corresponds to the conventional relation between renormalized and bare operators 
\begin{equation}
\mathcal{O}_\textrm{ph}^{(h,\gamma_5)\, bare}\ = \  Z_P^{-1}\,
\mathcal{O}_\textrm{ph}^{(h,\gamma_5)\, R}\ ;
\end{equation}
on the other hand, the conventional definition for the bare scalar densities, for which
\begin{equation}
\mathcal{O}_\textrm{ph}^{(h,1)\, bare,\,conv.}\ = \  Z_S^{-1}\,
\mathcal{O}_\textrm{ph}^{(h,1)\, R}\ 
\end{equation}
holds, is related to the definition (\ref{eq:bare_phys_corr}) by 
a finite renormalization
\begin{equation}
\mathcal{O}_\textrm{ph}^{(h,1)\, bare,\,conv.}\ = \  Z_P/Z_S\,
\mathcal{O}_\textrm{ph}^{(h,1)\, bare}\ .
\end{equation}
Of course, with $Z_P/Z_S$ at hand both definitions can be computed.

\begin{figure}[htb]
\begin{center}
\input{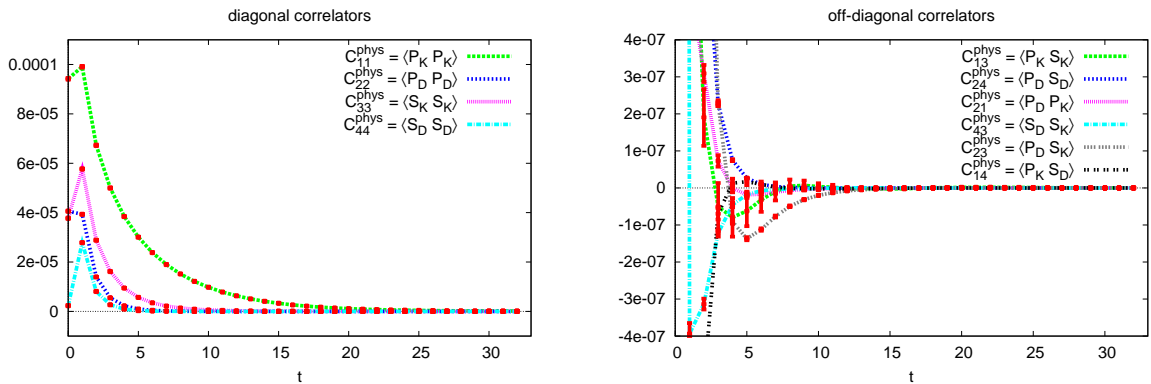}
\caption{\label{fig:phys_corr_B}Bare physical correlators for the 
ensemble B35.32 with Gaussian smeared operators.}
\end{center}
\end{figure}

Figure~\ref{fig:phys_corr_B} shows the diagonal and off-diagonal correlators of the bare 
physical operators for the ensemble B35.32 with Gaussian smearing, which are the ones 
used for the final computation of all masses. A general feature is that starting from time separation $t \gtapprox 5$ most of the off-diagonal elements become small and compatible with zero within statistical errors. An exception is the matrix element $\langle \mathcal{O}_\textrm{ph}^{(c,\gamma_5)}(\mathcal{O}_\textrm{ph}^{(s,1)})^\dagger \rangle$, which remains large and comparable in size with the two smallest diagonal elements in the $(c,-/+)$ sectors. At the moment we have no explanation for this observation. 

\begin{figure}[htb]
\begin{center}
\input{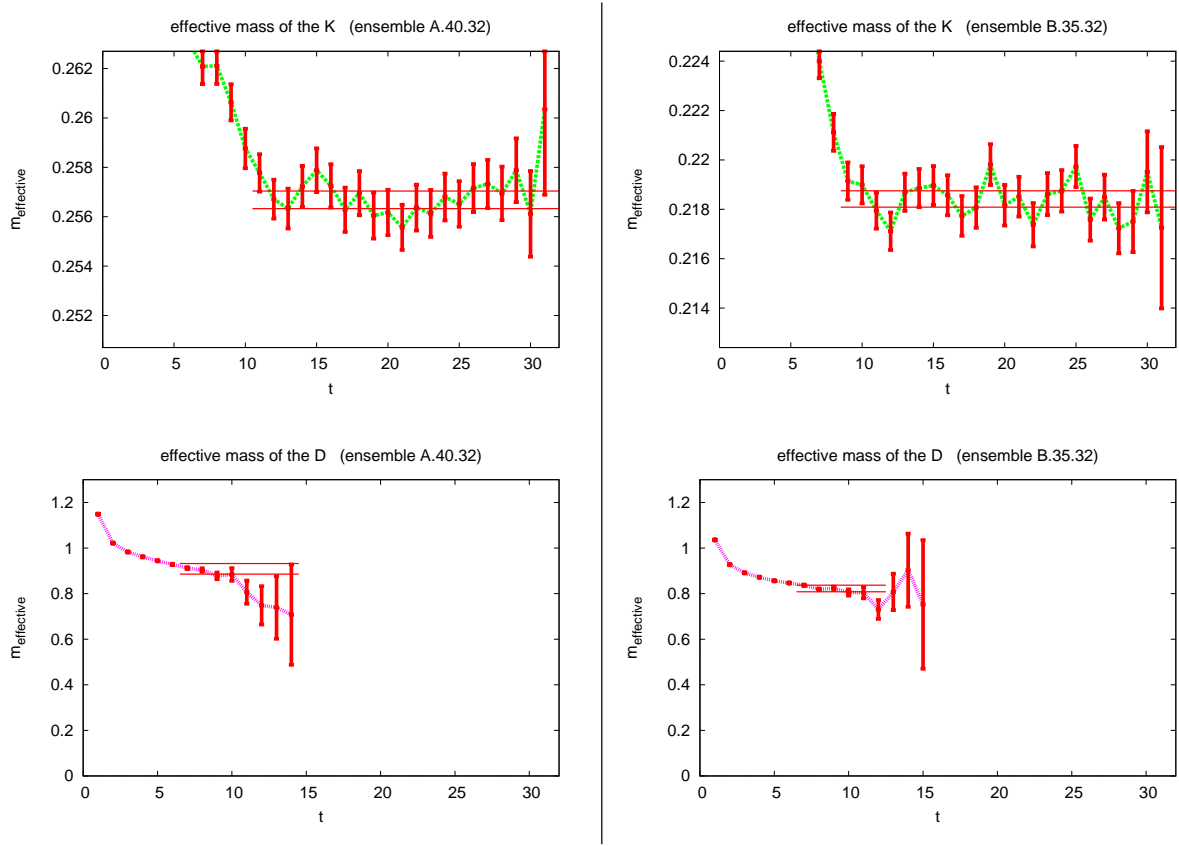}
\caption{\label{fig:eff_mass_PS} The effective masses in the pseudoscalar channel, with 
Gaussian smeared operators, for the ensembles A40.32 (left) and B35.32 (right). 
The error bands indicate the total error, statistical 
plus systematic. }
\end{center}
\end{figure}

It should also be noted that, following the arguments 
of~\cite{Frezzotti:2003ni,Frezzotti:2003xj}, $\mathcal{O}(a)$ improvement can only be 
expected for the diagonal elements of the physical correlation matrix. 
Since the twist angles and the ratio $Z_P/Z_S$ are obtained from conditions on the 
off-diagonal elements, one should a priori expect $\mathcal{O}(a)$ discretization 
errors for these quantities. However, it should be stressed that for 
physical quantities such 
as meson masses and decay constants, which are extracted from the diagonal matrix elements, 
$\mathcal{O}(a)$ improvement is at work.
The mass of the low-lying state in each of the four different channels can now be extracted 
by standard techniques from the diagonal correlator of the appropriate operator in
 (\ref{eq:bare_phys_corr}). The effective masses for the four channels and the two ensembles
 are reported in Figure~\ref{fig:eff_mass_PS}, for negative parity, and 
Figure~\ref{fig:eff_mass_S}, for positive parity. The final values for all
 masses are obtained by applying single-mass fits with a cosh function in the asymptotic 
regime. Also in this case the statistical error of the fitting parameters is determined by the linearization method of~\cite{WOLFF}. The starting time $t_{min}$ for the fits 
was chosen by requiring $\chi^2/\textrm{dof} \ltapprox 1$.

\begin{figure}[htb]
\begin{center}
\input{FIG008.pstex_t}
\caption{\label{fig:eff_mass_S} The effective masses in the scalar channel, with 
Gaussian smeared operators, for the ensembles A40.32 (left) and B35.32 (right).
The error bands indicate the total error, statistical 
plus systematic.}
\end{center}
\end{figure}
The plateaus for the charmed meson 
states are generally quite short, since the noise sets in early, typically around 
$t \gtapprox 11$. This is, however, expected. For those temporal separations the 
$D$ correlator is only a small fraction of the kaon correlator, as shown 
in the left panel of Figure~\ref{fig:phys_corr_B}. On the other hand, the $D$ correlator 
results from a linear combination of the correlators of the twisted basis 
$\chi$-field bilinears in eq.~(\ref{EQN001}), all dominated by the kaon. 
This means that the condition (\ref{eq:as_cond_me}) can only be fulfilled through a 
cancellation of large terms, one of the results being the comparably small $D$ correlator. 
The latter inherits the statistical fluctuations of the original bilinears and a large relative error is the consequence. As already stated many times in this paper, this is an inherent problem in our twisted mass setup, where the $D$ is actually a highly excited state in the 
mixed $(s/c,-/+)$ heavy-light meson sector.

In the case of the $D$ meson we attempt to estimate the systematic error produced by possible residual contributions of excited states and the influence of an 
unphysical mixing with the rather light $K_0^*$ state\footnote{The 
mixing with the kaon has been eliminated by construction.}. 
We apply a procedure analogous to the one of section~\ref{sec:method_2},
 and consider the spread of results by including all good fits 
(those with high significance) obtained by varying the fit interval 
$[t_\textrm{min}, t_\textrm{max}]$.
The resulting systematic error is much larger than the statistical one, and
 decreases on the finer lattice. This is reflected by the better quality of 
plateaus for the ensemble B35.32, as compared to A40.32, see Figure~\ref{fig:eff_mass_PS}.

The numerical results for all masses are listed in Table~\ref{tab:masses}. For $K^*_0$ 
different plateaus could be identified for the effective mass. In this case the value for 
each plateau is reported. It is unclear at this stage, whether this multi plateau 
behavior reflects the physical structure of QCD states in this sector, or is just a 
statistical effect, as also discussed at the end of section~\ref{SEC003}.

\begin{table}[htb]
\begin{center}
\renewcommand{\arraystretch}{1.0}
\begin{tabular}{|c|c|c|l|c|l|}
\hline
 \multicolumn{1}{|c|}{Ensemble} &
 \multicolumn{1}{|c|}{$am_K$} &
 \multicolumn{1}{|c|}{$t_1,t_2$} &
 \multicolumn{1}{|c|}{$am_{K_0^*}$} &
 \multicolumn{1}{|c|}{$am_D$} &
 \multicolumn{1}{|c|}{$am_{D_0^*}$} 
 \\
\hline
\hline
A40.32 &    0.25668(35)       &  7-8         &  0.452(8) &  0.909(4)(22) &  1.029(26)
 \\ \hline
\multicolumn{1}{|c|}{}   &              &  9-12     &  0.431(12)    & \multicolumn{1}{|c|}{} & \multicolumn{1}{|c|}{}   
 \\ \cline{3-4}
  \multicolumn{1}{|c|}{} &              &  14-32    &  0.37(5)      & \multicolumn{1}{|c|}{} & \multicolumn{1}{|c|}{}   
 \\ 
\cline{1-6}
\hline
B35.32 &    0.21842(33)        &  7-10       &  0.476(8) &  0.823(4)(14)  & 0.968(16) 
 \\ \hline
  \multicolumn{1}{|c|}{}  &             &  11-15    &  0.437(23) & \multicolumn{1}{|c|}{}  & \multicolumn{1}{|c|}{}     
 \\ \cline{3-4}
  \multicolumn{1}{|c|}{}  &             &  16-32    &  0.358(39) & \multicolumn{1}{|c|}{}  & \multicolumn{1}{|c|}{}    
 \\  \cline{1-6}
\end{tabular}

\caption{\label{tab:masses} Masses of the $K$, $K_0^\ast$, $D$ and $D_0^\ast$ 
mesons in lattice units, obtained with the parity and flavor restoration method, and using 
Gaussian smeared operators. The third row contains the temporal separations used for the 
determination of $m_{K_0^*}$.}

\end{center}
\end{table}

We conclude the illustration of this method by briefly discussing its generalization to the
case of $4\times n$ operators, the immediate application being the one considered in the previous section
with both local and smeared operators. The obvious route would just be to 
diagonalize each $4\times 4$ correlation sub-matrix with homogeneous composition
(e.g. local or smeared operators only) as we have done so far.  
As a result, the twist angles and the $Z_P/Z_S$ 
factors are obtained for each set; observe that the $Z_P/Z_S$ factors are heavily affected by the
smearing, which brings the former closer to one. Also the twist angles are expected to differ, due to 
different $\mathcal{O}(a)$ effects for local and smeared operators. Once these parameters are known, the physical 
correlation matrices with mixed local/smeared operators can be reconstructed, too.
However, this procedure is not expected to be optimal for the latter correlation matrices,
since the parameters are adjusted to optimize the correlation matrices with homogeneous 
composition. A better way would be to apply an independent diagonalization, with new 
parameters,
of the matrices with mixed local/smeared composition.
% ********************
% ********************
% ********************
% ********************
% ********************

\section{Conclusions}
\label{sec:concl}

\begin{figure}[htb]
\begin{center}
\input{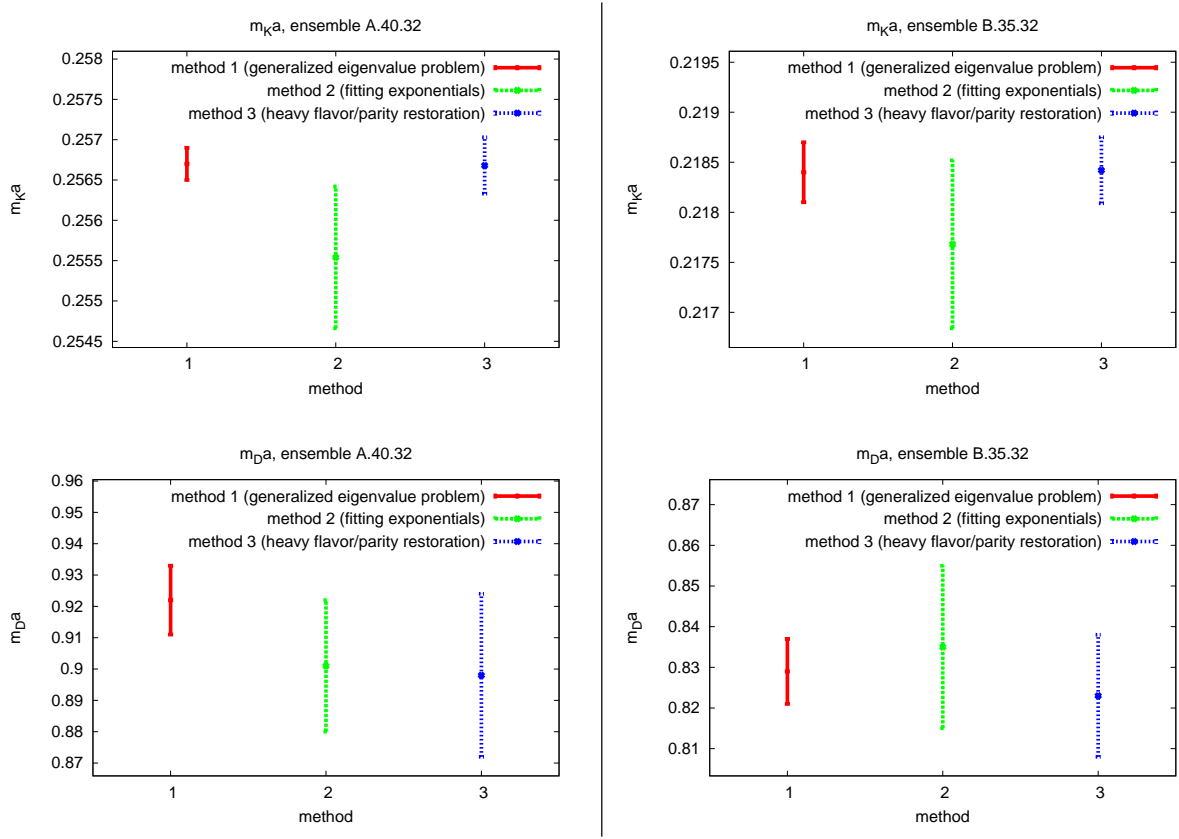}
\caption{\label{FIG009}Comparison of the results for $m_K$ (top) and $m_D$ (bottom)
obtained with the 
three methods exposed in this work, for both ensembles. The results for methods 1 to 3 
are shown from left to right.}
\end{center}
\end{figure}

We have proposed and compared three methods to determine $m_K$ and $m_D$ in 
$N_{\rm f} = 2+1+1$ twisted mass lattice QCD. The computation of these masses is less 
straightforward 
in this case, since parity and flavor are not good quantum numbers.
We have therefore explored strategies to extract the desired states and have developed 
three distinct methods all of which    
exploit the exponential fall-off of 
correlation matrices for suitably chosen heavy-light meson creation operators.  
Method~1 amounts to solving a generalized eigenvalue problem, 
method~2 is equivalent to fitting a linear superposition of exponentials and 
method~3 transforms the correlators to the physical basis by means of the twist rotation.
Results for $m_K$ and $m_D$ obtained with the three methods and for both ensembles 
investigated here are
summarized in Table~\ref{TAB009} and visualized in Figure~\ref{FIG009}.
\begin{table}[htb]
\begin{center}

\begin{tabular}{|c|c|c|c|}
\hline
 & & & \vspace{-0.40cm} \\
 & Method 1 & Method 2 & Method 3\vspace{-0.40cm} \\
 & & & \\
\hline
\multicolumn{4}{|c|}{\vspace{-0.40cm}} \\
\multicolumn{4}{|c|}{Ensemble A40.32} \\
\multicolumn{4}{|c|}{\vspace{-0.40cm}} \\
\hline
 & & & \vspace{-0.40cm} \\
$am_K $ & $0.2567(2)$ & $0.25554(88)$ & $0.25668(35)$  \\
$am_D $ & $0.922(11)$ & $0.901(21)$   & $0.909(22)$\vspace{-0.40cm} \\
 & & & \\
\hline
\multicolumn{4}{|c|}{\vspace{-0.40cm}} \\
\multicolumn{4}{|c|}{Ensemble B35.32} \\
\multicolumn{4}{|c|}{\vspace{-0.40cm}} \\
\hline
 & & & \vspace{-0.40cm} \\
$a m_K $ & $0.2184(3)$ & $0.21768(84)$ & $0.21842(33)$ \\
$a m_D$ & $0.829(8){\phantom 0}$  & $0.835(20)$   & $0.823(15)$\vspace{-0.40cm} \\
 & & & \\
\hline
\end{tabular}
\caption{\label{TAB009}Comparison of the results for $m_K$ and $m_D$ obtained with the 
three methods exposed in this work, for both ensembles.}
\end{center}
\end{table}
Since the kaon is the lightest state in the combined $(s/c,-/+)$ sector, 
the computation of its mass is rather simple and we obtain precise values for 
$m_K$ with errors $\ltapprox 0.4 \%$ including statistical and systematical 
uncertainties. Moreover, within these errors all three methods yield very 
compatible results which 
is very reassuring.

In contrast to $m_K$, the mass of the $D$ meson is difficult to determine, 
because in our twisted mass setup the $D$ meson is a highly excited state in the 
combined $(s/c,-/+)$ sector. However, also in this case our three methods yield results, 
which are in excellent agreement within the combined statistical and systematical errors,
whose relative magnitudes are $\ltapprox 2.5 \%$. 
Therefore, we are confident that we are able to obtain reliable estimates for $m_D$ 
without resolving all the low lying (multi particle) states below the $D$ meson. 
The latter would require to compute correlation matrices with a significantly larger 
operator basis and with extremely high statistical precision, an endeavor, 
which hardly seems to be feasible.
It is therefore very important that already with the much smaller correlator matrix employed 
here, 
%the $D$ meson mass can be extracted reliably and accurately.
one can obtain a satisfactory estimate of the $D$ meson mass.

The errors we obtain with our three methods differ by factors of around 
2 to 4, originating from the fact that the three methods estimate the systematic error in 
different ways. While method~2 (fitting exponentials) tends to yield the 
largest error, its procedure to determine the systematic error is also the 
most conservative: the error is computed from the spread of a large set of fit results 
corresponding to different fitting ranges. 
In contrast to that method~1 (solving a generalized eigenvalue problem) estimates 
the corresponding error by just taking two ``neighboring fitting ranges'' into account. 
Consequently, the total error is somewhat smaller.

We stress that as far as $K$ physics is concerned, our analysis shows
that this sector can be analyzed in the unitary setup without
problems. This provides a very good perspective to compute
corresponding decay constants and also the strange baryon spectrum in
the future. For charm physics, the situation is different and it will
be quite difficult to extract reliable physics results in the charm
sector from the unitary setup. Here, we plan to employ a mixed action
approach by using an Osterwalder-Seiler (OS) valence quark
action \cite{Frezzotti:2004wz}. This has the advantage \cite{Roberto} that there is no flavor mixing and that
the valence quarks stay as close as possible to the sea twisted mass
quarks, e.g.\ there is no need to re-tune $\kappa$ to realize maximal
twist.  The idea is to match the $K$ and $D$ meson masses between the
unitary setup and the valence OS quarks. After this matching step
further physical quantities such as decay constants will then be
computed with OS quarks. The matching condition will guarantee that in
the continuum limit we recover the situation of a unitary setup.  Of
course, it needs to be seen, whether discretization errors in this
strategy remain small. Investigations in this direction are in
progress.

With respect to the matching of $K$ and $D$ meson masses between the unitary setup and the valence OS quarks,
the outcome of our work in this paper is extremely important. The fact that we can compute 
% both masses with a high accuracy 
the $K$ meson with high accuracy and the $D$ meson with acceptable precision
in the unitary setup is a necessary prerequisite to allow 
for applying such a matching condition. 

% We finally want to mention that the ETMC has started a program to evaluate the 
% renormalization constants for our $N_{\rm f}=2+1+1$ setup. Once the relevant renormalization 
% constants will be available, we could also compute the renormalized strange and 
% charm quark masses directly by using eq.\ (\ref{eq:csmasses})
% and employ these as a matching condition \cite{Frezzotti:2004wz}. This will lead 
% to different tunings of the parameters in the valence sector and hence to different
% cut-off effects. Employing both matching conditions can therefore be used to 
% have independent computations for physical observables and will provide a most
% valuable cross-check of the continuum limit.  

Instead of matching the $K$ and $D$ meson masses in the sea and valence sectors,
one can directly match the renormalized strange and charm quark masses \cite{Frezzotti:2004wz}.
The latter can be determined in the sea sector by using eq.\ (\ref{eq:csmasses}).
Only the finite ratio $Z_P/Z_S$ is needed as an input for the matching. 
We have shown in this paper one possible way to determine this quantity,
which is specific for the twisted mass setup.
In compliance with the massless quark renormalization scheme, however, the 
extrapolated value of $Z_P/Z_S$ for four massless quarks is required. 
We mention here that the ETMC has started a dedicated program to evaluate 
the renormalization constants for our $N_{\rm f}=2+1+1$ setup in the massless quark limit.
Once the relevant renormalization constants will be available, 
this information will be used for an alternative tuning of 
the mass parameters in the valence sector. This can result in different values
of the valence quark masses with respect to the procedure relying on the hadron masses, 
and hence to different cut-off effects for the resulting mixed action theory.  
Employing both matching conditions can therefore be used to 
have independent computations for physical observables and will provide a most
valuable cross-check of the way this setup approaches the continuum limit.  

% ********************
% ********************
% ********************
% ********************
% ********************

\section*{Acknowledgments}

The computer time for this project was made available to us by the
John von Neumann-Institute for Computing (NIC) on the JUMP, Juropa and
Jugene systems in J\"ulich and apeNEXT system in Zeuthen, BG/P and
BG/L in Groningen, by BSC on Mare-Nostrum in Barcelona (www.bsc.es),
and by the computer resources made available by CNRS on the BlueGene
system at GENCI-IDRIS Grant 2009-052271 and CCIN2P3 in Lyon.  We thank
these computer centers and their staff for all technical advice and
help.

This work has been supported in part by the DFG
Sonder\-for\-schungs\-be\-reich/ Trans\-regio \\ SFB/TR9-03 and the EU
Integrated Infrastructure Initiative Hadron Physics (I3HP) under
contract RII3-CT-2004-506078. We also thank the DEISA Consortium
(co-funded by the EU, FP6 project 508830), for support within the
DEISA Extreme Computing Initiative (www.deisa.org).

% ********************
% ********************
% ********************
% ********************
% ********************

% ********************


\begin{thebibliography}{99}

\bibitem{Boucaud:2007uk}
  {\bf ETM} Collaboration, Ph.~Boucaud {\it et al.},
  ``Dynamical twisted mass fermions with light quarks,''
  Phys.\ Lett.\ B {\bf 650}, 304 (2007)
  [arXiv:hep-lat/0701012].
  %%CITATION = PHLTA,B650,304;%%

\bibitem{Boucaud:2008xu}
  {\bf ETM} Collaboration, Ph.~Boucaud {\it et al.},
  ``Dynamical twisted mass fermions with light quarks: simulation and analysis details,''
  Comput.\ Phys.\ Commun.\ {\bf 179}, 695 (2008)
  [arXiv:0803.0224 [hep-lat]].
  %%CITATION = CPHCB,179,695;%%

\bibitem{Baron:2009wt}
  {\bf ETM} Collaboration, R.~Baron {\it et al.},
  ``Light meson physics from maximally twisted mass lattice QCD,''
  arXiv:0911.5061 [hep-lat].
  %%CITATION = ARXIV:0911.5061;%%

\bibitem{Alexandrou:2008tn}
  {\bf ETM} Collaboration, C.~Alexandrou {\it et al.},
  ``Light baryon masses with dynamical twisted mass fermions,''
  Phys.\ Rev.\ D {\bf 78}, 014509 (2008)
  [arXiv:0803.3190 [hep-lat]].
  %%CITATION = PHRVA,D78,014509;%%

\bibitem{Alexandrou:2009qu}
  {\bf ETM} Collaboration, C.~Alexandrou {\it et al.},
  ``The low-lying baryon spectrum with two dynamical twisted mass fermions,''
  Phys.\ Rev.\ D {\bf 80}, 114503 (2009)
  [arXiv:0910.2419 [hep-lat]].
  %%CITATION = PHRVA,D80,114503;%%

\bibitem{Blossier:2007vv}
  {\bf ETM} Collaboration, B.~Blossier {\it et al.},
  ``Light quark masses and pseudoscalar decay constants from $N_{\rm f}=2$ lattice QCD with twisted mass fermions,''
  JHEP {\bf 0804}, 020 (2008)
  [arXiv:0709.4574 [hep-lat]].
  %%CITATION = JHEPA,0804,020;%%

\bibitem{Blossier:2009bx}
  {\bf ETM} Collaboration, B.~Blossier {\it et al.},
  ``Pseudoscalar decay constants of kaon and $D$-mesons from $N_{\rm f} = 2$ twisted mass lattice QCD,''
  JHEP {\bf 0907}, 043 (2009)
  [arXiv:0904.0954 [hep-lat]].
  %%CITATION = JHEPA,0907,043;%%

\bibitem{Bertone:2009bu}
  {\bf ETM} Collaboration, V.~Bertone {\it et al.},
  ``Kaon oscillations in the Standard Model and Beyond using $N_{\rm f}=2$ dynamical quarks,''
  PoS {\bf LAT2009 } (2009)  258.
  [arXiv:0910.4838 [hep-lat]].

\bibitem{Blossier:2009hg}
  {\bf ETM} Collaboration, B.~Blossier {\it et al.},
  ``A proposal for $B$-physics on current lattices,''
  arXiv:0909.3187 [hep-lat].
  %%CITATION = ARXIV:0909.3187;%%

\bibitem{Blossier:2009gd}
  {\bf ETM} Collaboration, B.~Blossier {\it et al.},
  ``$f_B$ and $f_{B_s}$ with maximally twisted Wilson fermions,''
  [arXiv:0911.3757 [hep-lat]].

\bibitem{Jansen:2008si}
  {\bf ETM} Collaboration, K.~Jansen, C.~Michael, A.~Shindler and M.~Wagner,
  ``The Static-light meson spectrum from twisted mass lattice QCD,''
  JHEP {\bf 0812}, 058 (2008)
  [arXiv:0810.1843 [hep-lat]].
  %%CITATION = JHEPA,0812,058;%%

\bibitem{:2010iv}
  {\bf ETM} Collaboration, C.~Michael, A.~Shindler and M.~Wagner,
  ``The continuum limit of the static-light meson spectrum,''
  arXiv:1004.4235 [hep-lat].
  %%CITATION = ARXIV:1004.4235;%%

\bibitem{Blossier:2009vy}
  {\bf ETM} Collaboration, B.~Blossier, M.~Wagner and O.~P\`ene,
  ``Lattice calculation of the Isgur-Wise functions $\tau_{1/2}$ and $\tau_{3/2}$ with dynamical quarks,''
  JHEP {\bf 0906}, 022 (2009)
  [arXiv:0903.2298 [hep-lat]].
  %%CITATION = JHEPA,0906,022;%%

\bibitem{Frezzotti:2008dr}
  {\bf ETM} Collaboration, R.~Frezzotti, V.~Lubicz and S.~Simula,
  ``Electromagnetic form factor of the pion from twisted-mass lattice QCD at $N_{\rm f}=2$,''
  Phys.\ Rev.\  D {\bf 79}, 074506 (2009)
  [arXiv:0812.4042 [hep-lat]].
  %%CITATION = PHRVA,D79,074506;%%

\bibitem{Lubicz:2009ht}
  {\bf ETM} Collaboration, V.~Lubicz, F.~Mescia, S.~Simula {\it et al.},
  ``$K \to \pi \ell \nu$ semileptonic form factors from two-flavor lattice QCD,''
  Phys.\ Rev.\  {\bf D80 } (2009)  111502.
  [arXiv:0906.4728 [hep-lat]].

%\cite{DiVita:2009by}
\bibitem{DiVita:2009by}
  {\bf ETM} Collaboration, S.~Di Vita {\it et al.},
  ``Vector and scalar form factors for $K$- and $D$-meson semileptonic
  decays from twisted mass fermions with $N_{\rm f} = 2$,''
  PoS {\bf LAT2009 } (2009)  257.
  [arXiv:0910.4845 [hep-lat]].

\bibitem{Alexandrou:2008rp}
  {\bf ETM} Collaboration, C.~Alexandrou {\it et al.},
  ``Nucleon form factors with dynamical twisted mass fermions,''
  PoS {\bf LATTICE2008}, 139 (2008)
  [arXiv:0811.0724 [hep-lat]].
  %%CITATION = POSCI,LATTICE2008,139;%%

\bibitem{Baron:2007ti}
  {\bf ETM} Collaboration, R.~Baron, S.~Capitani, J.~Carbonell, K.~Jansen, Z.~Liu, O.~Pene and C.~Urbach,
  ``Moments of meson distribution functions with dynamical twisted mass fermions,''
  PoS {\bf LAT2007}, 153 (2007)
  [arXiv:0710.1580 [hep-lat]].
  %%CITATION = POSCI,LAT2007,153;%%

\bibitem{Michael:2007vn}
  {\bf ETM} Collaboration, C.~Michael and C.~Urbach,
  ``Neutral mesons and disconnected diagrams in twisted mass QCD,''
  PoS {\bf LAT2007}, 122 (2007)
  [arXiv:0709.4564 [hep-lat]].
  %%CITATION = POSCI,LAT2007,122;%%

\bibitem{Jansen:2008wv}
  {\bf ETM} Collaboration, K.~Jansen, C.~Michael and C.~Urbach,
  ``The $\eta'$ meson from lattice QCD,''
  Eur.\ Phys.\ J.\  C {\bf 58}, 261 (2008)
  [arXiv:0804.3871 [hep-lat]].
  %%CITATION = EPHJA,C58,261;%%

\bibitem{McNeile:2009mx}
  {\bf ETM} Collaboration, C.~McNeile, C.~Michael and C.~Urbach,
  ``The $\omega$-$\rho$ meson mass splitting and mixing from lattice QCD,''
  Phys.\ Lett.\  B {\bf 674}, 286 (2009)
  [arXiv:0902.3897 [hep-lat]].
  %%CITATION = PHLTA,B674,286;%%

\bibitem{Renner:2009by}
  {\bf ETM} Collaboration, D.~B.~Renner and X.~Feng,
  ``Hadronic contribution to $g-2$ from twisted mass fermions,''
  PoS {\bf LATTICE2008}, 129 (2008)
  [arXiv:0902.2796 [hep-lat]].
  %%CITATION = POSCI,LATTICE2008,129;%%

\bibitem{Feng:2009ij}
  {\bf ETM} Collaboration, X.~Feng, K.~Jansen and D.~B.~Renner,
  ``The $\pi^+$ $\pi^+$ scattering length from maximally twisted mass lattice QCD,''
  Phys.\ Lett.\  B {\bf 684}, 268 (2010)
  [arXiv:0909.3255 [hep-lat]].
  %%CITATION = PHLTA,B684,268;%%

\bibitem{Feng:2009ck}
  {\bf ETM} Collaboration, X.~Feng, K.~Jansen and D.~B.~Renner,
  ``Scattering from finite size methods in lattice QCD,''
  arXiv:0910.4871 [hep-lat].
  %%CITATION = ARXIV:0910.4871;%%

\bibitem{Constantinou:2010gr}
  {\bf ETM} Collaboration, M.~Constantinou {\it et al.},
  ``Non-perturbative renormalization of quark bilinear operators with
  $N_{\rm f}=2$
  (tmQCD) Wilson fermions and the tree-level improved gauge action,''
  arXiv:1004.1115 [hep-lat].
  %%CITATION = ARXIV:1004.1115;%%

\bibitem{Cichy:2008gk}
  K.~Cichy, J.~Gonzalez Lopez, K.~Jansen, A.~Kujawa and A.~Shindler,
  ``Twisted mass, overlap and Creutz fermions: cut-off effects at tree-level of perturbation theory,''
  Nucl.\ Phys.\  B {\bf 800}, 94 (2008)
  [arXiv:0802.3637 [hep-lat]].
  %%CITATION = NUPHA,B800,94;%%

\bibitem{Jansen:2005gf}
  {\bf \xlf} Collaboration, K.~Jansen, M.~Papinutto, A.~Shindler, C.~Urbach and I.~Wetzorke,
  ``Light quarks with twisted mass fermions,''
  Phys.\ Lett.\  B {\bf 619}, 184 (2005)
  [arXiv:hep-lat/0503031].
  %%CITATION = PHLTA,B619,184;%%

\bibitem{Jansen:2005kk}
  {\bf \xlf} Collaboration, K.~Jansen, M.~Papinutto, A.~Shindler, C.~Urbach and I.~Wetzorke,
  ``Quenched scaling of Wilson twisted mass fermions,''
  JHEP {\bf 0509}, 071 (2005)
  [arXiv:hep-lat/0507010].
  %%CITATION = JHEPA,0509,071;%%

\bibitem{Abdel-Rehim:2005gz}
  A.~M.~Abdel-Rehim, R.~Lewis and R.~M.~Woloshyn,
  ``Spectrum of quenched twisted mass lattice QCD at maximal twist,''
  Phys.\ Rev.\  D {\bf 71}, 094505 (2005)
  [arXiv:hep-lat/0503007].
  %%CITATION = PHRVA,D71,094505;%%

\bibitem{Urbach:2007rt}
  {\bf ETM} Collaboration, C.~Urbach,
  ``Lattice QCD with two light Wilson quarks and maximally twisted mass,''
  PoS {\bf LAT2007 } (2007)  022
  [arXiv:0710.1517 [hep-lat]].

\bibitem{Dimopoulos:2007qy}
  {\bf ETM} Collaboration, P.~Dimopoulos, R.~Frezzotti, G.~Herdoiza, C.~Urbach and U.~Wenger,
  ``Scaling and low energy constants in lattice QCD with $N_{\rm f}=2$ maximally
  twisted Wilson quarks,''
  PoS {\bf LAT2007} (2007) 102
  [arXiv:0710.2498 [hep-lat]].
  %%CITATION = POSCI,LAT2007,102;%%

\bibitem{Frezzotti:2005gi}
  R.~Frezzotti, G.~Martinelli, M.~Papinutto and G.~C.~Rossi,
  ``Reducing cutoff effects in maximally twisted lattice QCD close to the chiral limit,''
  JHEP {\bf 0604} 038 (2006)
  [arXiv:hep-lat/0503034].
  %%CITATION = JHEPA,0604,038;%%

\bibitem{Dimopoulos:2009qv}
  {\bf ETM} Collaboration, P.~Dimopoulos, R.~Frezzotti, C.~Michael, G.~C.~Rossi and C.~Urbach,
  ``$\mathcal{O}(a^2)$ cutoff effects in lattice Wilson fermion simulations,''
  Phys.\ Rev.\  D {\bf 81}, 034509 (2010)
  [arXiv:0908.0451 [hep-lat]].
  %%CITATION = PHRVA,D81,034509;%%

\bibitem{Frezzotti:2004wz}
  R.~Frezzotti and G.~C.~Rossi,
  ``Chirally improving Wilson fermions. II: Four-quark operators,''
  JHEP {\bf 0410} (2004) 070
  [arXiv:hep-lat/0407002].
  %%CITATION = JHEPA,0410,070;%%

\bibitem{Chiarappa:2006ae}
  T.~Chiarappa {\it et al.},
  ``Numerical simulation of QCD with $u$, $d$, $s$ and $c$ quarks in the twisted-mass Wilson formulation,''
  Eur.\ Phys.\ J.\  C {\bf 50}, 373 (2007)
  [arXiv:hep-lat/0606011].
  %%CITATION = EPHJA,C50,373;%%

\bibitem{Baron:2008xa}
  {\bf ETM} Collaboration, R.~Baron {\it et al.},
  ``Status of ETMC simulations with $N_{\rm f} = 2+1+1$ twisted mass fermions,''
  PoS {\bf LATTICE2008}, 094 (2008)
  [arXiv:0810.3807 [hep-lat]].
  %%CITATION = POSCI,LATTICE2008,094;%%

\bibitem{Baron:2009zq}
  {\bf ETM} Collaboration, R.~Baron {\it et al.},
  ``First results of ETMC simulations with $N_{\rm f} = 2+1+1$ maximally twisted mass fermions,''
  arXiv:0911.5244 [hep-lat].
  %%CITATION = ARXIV:0911.5244;%%

\bibitem{Baron:2010LL} {\bf ETM} Collaboration, R.~Baron {\it et al.},
  ``Light hadrons from lattice QCD with light ($u,d$), strange and
  charm dynamical quarks'', arXiv:1004.5284 [hep-lat].

\bibitem{Bazavov:2009jc}
  {\bf MILC} Collaboration, A.~Bazavov {\it et al.},
  ``HISQ action in dynamical simulations,''
  PoS {\bf LATTICE2008} (2008) 033
  [arXiv:0903.0874 [hep-lat]].
  %%CITATION = POSCI,LATTICE2008,033;%%

\bibitem{Bazavov:2009wm}
 {\bf MILC} Collaboration, A.~Bazavov {\it et al.},
  ``Progress on four flavor QCD with the HISQ action,''
  PoS {\bf LAT2009} (2009) 123
  [arXiv:0911.0869 [hep-lat]].
  %%CITATION = POSCI,LAT2009,123;%%

\bibitem{Bazavov:2010ru}
 {\bf MILC} Collaboration, A.~Bazavov {\it et al.},
  ``Scaling studies of QCD with the dynamical HISQ action,''
  [arXiv:1004.0342 [hep-lat]].
  %%CITATION = ARXIV:1004.0342;%%

\bibitem{Pena:2004gb}
  C.~Pena, S.~Sint and A.~Vladikas,
  ``Twisted mass QCD and lattice approaches to the Delta(I) = 1/2 rule,''
  JHEP {\bf 0409} (2004) 069
  [arXiv:hep-lat/0405028].
  %%CITATION = JHEPA,0409,069;%%

\bibitem{AbdelRehim:2006ve}
  A.~M.~Abdel-Rehim, R.~Lewis, R.~M.~Woloshyn and J.~M.~S.~Wu,
  ``Strange quarks in quenched twisted mass lattice QCD,''
  Phys.\ Rev.\  D {\bf 74} (2006) 014507
  [arXiv:hep-lat/0601036].
  %%CITATION = PHRVA,D74,014507;%%

\bibitem{Iwasaki:1985we}
  Y.~Iwasaki,
  ``Renormalization group analysis of lattice theories and improved lattice action: two-dimensional non-linear $\mathcal{O}(N)$ sigma model,''
  Nucl.\ Phys.\ B {\bf 258}, 141 (1985).
  %%CITATION = NUPHA,B258,141;%%

\bibitem{Frezzotti:2000nk}
  {\bf ALPHA} Collaboration, R.~Frezzotti, P.~A.~Grassi, S.~Sint and P.~Weisz,
  ``Lattice QCD with a chirally twisted mass term,''
  JHEP {\bf 0108}, 058 (2001)
  [arXiv:hep-lat/0101001].
  %%CITATION = JHEPA,0108,058;%%

\bibitem{Frezzotti:2003xj}
  R.~Frezzotti and G.~C.~Rossi,
  ``Twisted-mass lattice QCD with mass non-degenerate quarks,''
  Nucl.\ Phys.\ Proc.\ Suppl.\  {\bf 128} (2004) 193
  [arXiv:hep-lat/0311008].
  %%CITATION = NUPHZ,128,193;%%

\bibitem{Frezzotti:2003ni}
  R.~Frezzotti and G.~C.~Rossi,
  ``Chirally improving Wilson fermions. I: $\mathcal{O}(a)$ improvement,''
  JHEP {\bf 0408}, 007 (2004)
  [arXiv:hep-lat/0306014].
  %%CITATION = JHEPA,0408,007;%%

\bibitem{PDG}
  C.\ Amsler {\it et al.} [Particle Data Group],
  Phys.\ Lett\. B {\bf 667}, 1 (2008) and 2009 partial update for the 2010 edition.

\bibitem{Blossier:2009kd}
  {\bf ALPHA} Collaboration, B.~Blossier, M.~Della Morte, G.~von Hippel, T.~Mendes and R.~Sommer,
  ``On the generalized eigenvalue method for energies and matrix elements in lattice field theory,''
  JHEP {\bf 0904}, 094 (2009)
  [arXiv:0902.1265 [hep-lat]].
  %%CITATION = JHEPA,0904,094;%%

\bibitem{WOLFF}
  {\bf ALPHA} Collaboration, U.~Wolff,
  ``Monte Carlo errors with less errors,''
  Comput.\ Phys.\ Commun.\ {\bf 156}, 143 (2004)
  [Erratum-ibid.\ {\bf 176}, 383 (2007)]
  [arXiv:hep-lat/0306017].
  %%CITATION = CPHCB,156,143;%%

\bibitem{MICHAEL}
  C.~Michael,
  ``Fitting correlated data,''
  Phys.\ Rev.\ D {\bf 49}, 2616 (1994)
  [arXiv:hep-lat/9310026].
  %%CITATION = PHRVA,D49,2616;%%

\bibitem{Michael:1994sz}
  C.~Michael and A.~McKerrell,
  ``Fitting Correlated Hadron Mass Spectrum Data,''
  Phys.\ Rev.\ D {\bf 51}, 3745 (1995)
  [arXiv:hep-lat/9412087].
  %%CITATION = PHRVA,D51,3745;%%

\bibitem{Farchioni:2004ma}
  F.~Farchioni {\it et al.},
  ``Exploring the phase structure of lattice QCD with twisted mass quarks,''
  Nucl.\ Phys.\ Proc.\ Suppl.\ {\bf 140}, 240 (2005)
  [arXiv:hep-lat/0409098].
  %%CITATION = NUPHZ,140,240;%%

\bibitem{Farchioni:2004fs}
  F.~Farchioni {\it et al.},
  ``The phase structure of lattice QCD with Wilson quarks and renormalization group improved gluons,''
  Eur.\ Phys.\ J.\ C {\bf 42}, 73 (2005)
  [arXiv:hep-lat/0410031].
  %%CITATION = EPHJA,C42,73;%%

\bibitem{Farchioni:2005bh}
  F.~Farchioni {\it et al.},
  ``Numerical simulations with two flavours of twisted-mass Wilson quarks and DBW2 gauge action,''
  Eur.\ Phys.\ J.\ C {\bf 47}, 453 (2006)
  [arXiv:hep-lat/0512017].
  %%CITATION = EPHJA,C47,453;%%

\bibitem{Roberto}
  R.~Frezzotti,
  private communication (2008).

\end{thebibliography}
\end{document}